%% file: SUSY_Anomaly_ArXiv2.tex
\documentclass[12pt]{article}
\pdfoutput=1

\usepackage{amsmath,amssymb,amsfonts,amstext,epsfig,cite,setspace,bigstrut,framed}
\usepackage[all]{xy}
\usepackage{color,xcolor,soul}
\usepackage{pifont}
\usepackage{latexsym}
\usepackage{graphics,graphicx}
\usepackage{hyperref}
\usepackage[bbgreekl]{mathbbol}
\usepackage[utf8]{inputenc}
\usepackage{fancyhdr}
\usepackage{lastpage}
\usepackage{etoolbox}
\usepackage{setspace}
\usepackage{titlesec}
\usepackage{cancel}
\usepackage{comment}
\usepackage[framemethod=tikz]{mdframed}
\usepackage{empheq}
\usepackage{tikz}
\usepackage{tcolorbox}
\tcbuselibrary{theorems}
\usepackage[font=footnotesize,labelfont=bf,justification=centerlast,width=.94\textwidth]{caption}

\makeatletter \@addtoreset{equation}{section} \makeatother
\renewcommand{\theequation}{\thesection.\arabic{equation}}

\def\rmd{\rm d}

\def\bo{{\beta_1}}

\include{macros}

\setul{0.5ex}{0.3ex}
    \definecolor{Red}{rgb}{1,0.0,0.0}
    \setulcolor{Red}

\newmdenv[%
middlelinecolor=gray!30!,
middlelinewidth=1pt,
backgroundcolor=gray!10!,
roundcorner=3pt
]{myidentity}

\mdfdefinestyle{mystyle}{%
middlelinecolor=gray!80!white,
middlelinewidth=1pt,
backgroundcolor=blue!5!white,
roundcorner=10pt,
subtitlebelowline=true,
frametitlefont={\normalfont\bfseries\sffamily\color{gray!90!white}},
innertopmargin=.1cm,
innerbottommargin=.3cm,
}

\def\bebx #1 \eebx{\begin{empheq}[box={\tcbhighmath[colframe=gray!90!white,colback=white]}]{align} #1 \end{empheq}}

\def\bbxd#1\ebxd{\begin{myidentity} \vskip -.4cm #1 \end{myidentity}\vskip-.2cm}

\def\bframe#1{\begin{mdframed}[style=mystyle,frametitle={#1}]}
\def\eframe{\end{mdframed}}

\newcommand{\bc}{{\bf c}}
\newcommand{\bq}{{\bf q}}
\newcommand{\bs}{{\bf s}}
\newcommand{\bx}{{\bf x}}

\newcommand{\cJ}{{\cal J}}

\newcommand{\cW}{{\cal W}}
\newcommand{\cN}{{\cal N}}
\newcommand{\cR}{{\cal R}}
\newcommand{\cG}{{\cal G}}

\newcommand{\cL}{{\cal L}}

\newcommand{\cF}{{\cal F}}
\newcommand{\cA}{{\cal A}}
\newcommand{\cX}{{\cal X}}

\newcommand{\bi}{\begin{itemize}}
\newcommand{\ei}{\end{itemize}}
\newcommand{\beq}{\begin{equation}}
\newcommand{\eeq}{\end{equation}}

\def\theequation{\thesection.\arabic{equation}}

\renewcommand{\thefootnote}{\fnsymbol{footnote}}

\begin{document}
\begin{titlepage}
\addtocounter{page}{-2}

\vfill
\begin{flushright}
{\tt\normalsize KIAS-P21011}\\

\end{flushright}
\vfill

\begin{center}
{\Large\bf Anomalies and Supersymmetry}

\vskip 1.5cm
Ruben Minasian$^1$\footnote{\tt ruben.minasian@ipht.fr}, Ioannis Papadimitriou$^2$\footnote{\tt ioannis@kias.re.kr}, Piljin Yi$^2$\footnote{\tt piljin@kias.re.kr}

\vskip 5mm

{\it $^1$Institut de Physique Théorique, Université Paris Saclay, CNRS, CEA, F-91191, Gif-sur-Yvette, France}
\vskip.2cm
{\it $^2$School of Physics, Korea Institute for Advanced Study, Seoul 02455, Korea}

\end{center}
\vfill

\begin{abstract}
We revisit quantum field theory anomalies, emphasizing the interplay
with diffeomorphisms and supersymmetry. The Ward identities of the latter
induce Noether currents of all continuous symmetries, and we point out
how these consistent currents are replaced by their covariant form through
the appearance of the Bardeen-Zumino currents, which play a central role
in our study. For supersymmetry Ward identities, two systematic methods
for solving the Wess-Zumino consistency conditions are discussed: anomaly inflow and
anomaly descent. The simplest inflows are from supersymmetric Chern-Simons
actions in one dimension higher, which are used to supersymmetrize flavor
anomalies in $d=4$ and, for $d=2$ $\cn=(p,q)$, flavor anomalies
with $p,q\leq 3$ and Lorentz-Weyl anomalies with $p,q\leq 6$. Finally,
we extend the BRST algebra and the subsequent descent, a necessity for the
diffeomorphism anomaly in retrospect. The same modification computes
the supersymmetrized anomalies, and determines the above Chern-Simons
actions when these exist.

\end{abstract}

\vfill
\end{titlepage}

\tableofcontents
\addtocontents{toc}{\protect\setcounter{tocdepth}{2}}
\renewcommand{\thefootnote}{\arabic{footnote}}
\setcounter{footnote}{0}

\section{Introduction and Summary of Results}
\label{sec:intro}

Perturbative anomalies \cite{Adler:1969gk, Bardeen:1969md, Bell:1969ts}
are well understood as a failure of the path integral measure to respect
the symmetry in question \cite{Fujikawa:1979ay}. Perhaps the most succinct
way to compute and phrase the anomaly is by gauging the would-be symmetry,
and consider the symmetry transformation of the effective action after
integrating out the chiral fields responsible for the anomaly. A most
comprehensive computation of this kind was given in Ref.~\cite{AlvarezGaume:1983ig}.
Further subtleties, such as the distinction between consistent and covariant currents, 
diffeomorphisms and Lorentz transformations, were addressed by Ref.~\cite{Bardeen:1984pm}.

Through these developments, we are accustomed to treating gravitational or diffeomorphism
anomalies on an equal footing with those associated with internal
symmetries, such as gauge or flavor symmetries. More often than not,
instead of the actual diffeomorphism anomaly we compute what is better
referred to as a Lorentz anomaly under an $SO(d)$ gauge rotation of
the spin connection $\o$,
\be\label{Lorentz}
\delta \om^a_{\;\;b} = \tx dL^a_{\;\;b} + \om^a_{\;\;c}L^c_{\;\;b}- L^a_{\;\;c}\om^c_{\;\;b}\,.
\ee
Although the equivalence of the gravitational and Lorentz anomalies has been
established a long time ago \cite{Bardeen:1984pm}, their precise relation is
far from obvious, given that diffeomorphisms involve both $SO(d)$ rotations
and translations.

Diffeomorphisms act via the Lie derivative along a
vector field, say $\xi$. They induce a Lorentz transformation, parameterized
by $L$ in \eqref{Lorentz}, via $\nabla \xi$, but also involve a translational
shift of $\om$. As we review in Section \ref{sec:diff}, despite this difference,
the diffeomorphism anomaly is still computable by a $GL(d)$ anomaly descent
on the Christoffel symbol, $\Gamma$, suitably elevated to a
1-form connection. This helps establish the equivalence between these two
anomalies, which justifies the usual focus on the Lorentz anomaly, especially
when one is only interested in the question of anomaly cancellation.

When it comes to the anomalous Ward identities with non-vanishing
anomalies, however, we must be more attentive to such differences.
After all, Lorentz rotations and diffeomorphisms are two
different operations. The diffeomorphism Ward identity
involves not only the divergence of the energy-momentum tensor,
but also all other symmetry currents, since the Lie derivative
acts on currents universally, which can be in turn converted
to a transformation of the associated (external) gauge fields via the
path integral. Of course, the converse is not true, since
gauge/flavor rotations do not affect the energy-momentum tensor.

Two distinct anomalous currents are often discussed: consistent
and covariant \cite{Bardeen:1984pm}.
The former arise from a direct variation of the effective action.
The difference between the currents is given by the Bardeen-Zumino (BZ) current,
a local quantity built out of the external gauge fields and
determined entirely by the anomaly descent. In the literature,
its use is mostly restricted to helping clarify the relation
between different computations of anomalies. However, as we will see,
it plays a central role in our discussion.

Since the consistent current is the one that couples to the gauge
field, it is often considered as more physical, while its covariant
counterpart has played a relatively minor role. On the other hand,
it is not difficult to see that the diffeomorphism  Ward identity is
invariant under such gauge/flavor symmetries, and so are the energy-momentum
tensor and the diffeomorphism anomaly. So, even though the consistent
currents of such internal symmetries  appear naturally in the path
integral derivation of the diffeomorphism  Ward identity, they must
somehow be replaced by their covariant counterparts to ensure the
gauge/flavor invariance of the diffeomorphism Ward identity.
The way the BZ terms that relate the consistent and covariant
gauge/flavor currents arise from a rearrangement of the diffeomorphism
Ward identity is the main subject of Section \ref{sec:diff}.

Another important Ward identity which generally involves other
symmetry currents is that of supersymmetry. Since the square of
supersymmetry generates translations or diffeomorphisms, it is
not surprising that the two suffer from some common issues.\footnote{$R$-symmetry
share features with both gauge/flavor symmetries and supersymmetry/diffeomorphisms.
For rudimentary discussions, it may be treated as one of the former but not
so when we get down to details of how it enters the Ward identities in question.
We will try to indicate such differences explicitly as it becomes necessary in later
parts of this note.}  Supersymmetry acts on  all gauge fields, external or internal,
and as a consequence the conservation of the supercurrent, as in the
diffeomorphism Ward identity, involves all gauge/flavor currents present.
In a supersymmetric theory with anomalous internal symmetries, one
naturally elevates the accompanying gauge fields to vector supermultiplets.
As a result, the supersymmetry Ward identity involves all operators in the corresponding current supermultiplets. In Section \ref{sec:susy} we
demonstrate that, as with diffeomorphisms, the BZ current terms are generated and all consistent currents in the supersymmetry Ward identity are eventually replaced by their covariant
counterparts.

However, there are two related important differences between diffeomorphisms
and supersymmetry in this respect. Firstly, the presence of gauge/flavor
anomalies does not violate the diffeomorphism or Lorentz invariance of the
effective action, nor does it generate unexpected contributions to the
diffeomorphism Ward identity. In this case, the covariantization of the current occurs
via rearrangement of existing terms. As we discuss in Section \ref{sec:susy},
this is not the case for supersymmetry in the Wess-Zumino (WZ) gauge for
the gauge multiplet. In this gauge, which is sometimes unavoidable, any
perturbative anomaly necessarily leads to non-invariance of the effective
action under rigid supersymmetry, an observation dating back to the 1980's \cite{Piguet:1980fa,Itoyama:1985qi,Guadagnini:1985ea,Itoyama:1985ni,Girardi:1985hf,Hwang:1985tm,Altevogt:1987qe,
Kaiser:1988zg,Altevogt:1987fx,Altevogt:1989fw,Baulieu:2006gx,Baulieu:2008id,Bzowski:2020tue} and often referred to as a ``supersymmetry anomaly".
An analogous observation in the presence of a gravitational anomaly
in two dimensions was first made in Refs.~\cite{Tanii:1985wy,Howe:1985uy}. Locally supersymmetric contributions to gauge/flavor and $R$-symmetry anomalies were pointed out in Ref.~\cite{Gates:1981yc,deWit:1985bn} and computed explicitly more recently in Refs.~\cite{Papadimitriou:2017kzw,An:2017ihs,Papadimitriou:2019gel,Papadimitriou:2019yug,An:2019zok,Kuzenko:2019vvi,Katsianis:2019hhg,Katsianis:2020hzd,Nakagawa:2021wqh}. See also Refs.~\cite{Bonora:1985ug,Brandt:1993vd,Brandt:1996au} for a classification of supersymmetrized supergravity anomalies.

The second difference between supersymmetry and diffeomorphisms is that, the
former does not have an independent solution to the WZ consistency
conditions. One might naively think that the supersymmetry Ward identity must
be thus simpler than that of diffeomorphisms, yet the situation is actually
the opposite. In the supersymmetry Ward identities, the leading ``anomalous" term
due to any other types of anomalies, gauge/flavor, R-symmetry, and diffeomorphisms,
is given by the BZ current contributions that covariantize the consistent currents thereof. Such a noncovariant term could have been expected on general grounds, given that the supersymmetry  Ward identity should be invariant under these other symmetries. However, this shift by the BZ current alone does not solve all WZ consistency conditions, and so on the right hand side of the Ward identity one encounters additional invariant pieces, involving the superpartners of the gauge fields.

These two differences are also behind the fact that the mechanism for the appearance of
the BZ currents for gauge/flavor (and $R$-symmetries) in the supersymmetry and
diffeomorphism Ward identities is not exactly the same. As we explain in detail
in Sections \ref{sec:diff} and \ref{sec:susy}, the covariantization of the currents
in the latter arises because the diffeomorphism Ward identity contains a linear
combination of the anomalous Ward identities for all gauge/flavor symmetries (and $R$-symmetries). In both cases, nevertheless, the covariantization of the currents is facilitated by the fundamental relation between the BZ current and the corresponding anomaly.

A superspace description, when it exists, ensures that the supersymmetry
invariance of the effective action may be restored by extending the multiplet of
currents to a larger one \cite{Kuzenko:2019vvi,Katsianis:2020hzd,Bzowski:2020tue,Nakagawa:2021wqh}.
This however does not change the actual content of the Ward identity; it merely gives
different names to the same local terms. Furthermore the WZ gauge for the gauge fields
is often a necessity, as for theories with extended supersymmetries. The WZ gauge
is often appropriate and sometimes unavoidable framework for computing (refined)
physical observables in supersymmetric theories using, e.g., supersymmetric
localization. It was in this context that the supersymmetrized form of the
gauge/flavor and, in particular, $R$-symmetry anomalies were recently rediscovered \cite{Papadimitriou:2017kzw}
and their consequences for supersymmetric partition functions explored \cite{Closset:2019ucb}.
We will refer to these phenomena as ``supersymmetrized anomaly" or ``supersymmetric
completion of anomaly" in most of this note.\footnote{The latter should not be
mistaken to imply that the anomalies under consideration are part of an anomaly multiplet.}

Supersymmetrization of various anomalies is a direct consequence of the WZ
consistency conditions in supersymmetric theories and ought to fit naturally
within established methods for solving these, such as the descent
formalism\cite{Atiyah:1984tf, Faddeev:1985iz, Manes:1985df,Bardeen:1984pm} and,
more physically, the anomaly inflow \cite{Callan:1984sa}.

Anomaly inflow provides a cancellation mechanism for gauge/flavor, $R$-symmetry,
and gravitational anomalies via, for example, the symmetry transformation of
a Chern-Simons action in one dimension higher. It is only natural that
a supersymmetric Chern-Simons action, if it exists, cancels all components of
the supersymmetrized anomalies. In Section \ref{sec:inflow} we show that this
is indeed the case and, turning the argument around, we use anomaly inflow as
a powerful tool to compute supersymmetric anomalies in theories with extended rigid or local supersymmetry.

Anomaly descent provides a more abstract approach to the solution of the
WZ consistency conditions. An interesting question in this context is how
the interplay between internal and spacetime symmetries is reflected in
the descent procedure. The standard anomaly descent applies to gauge/flavor
symmetries, as well as local Lorentz and $R$-symmetry, but it becomes less
obvious already for diffeomorphisms, given that these generate translations
as well as $GL(d)$ rotations. Supersymmetry adds a further layer of complication.
Remarkably, as we show in Section \ref{sec:descent}, diffeomorphisms and
supersymmetry can be accommodated by a single generalization of the standard descent procedure.

In both anomaly inflow and descent, Chern-Simons actions in one dimension
higher have a prominent role. In anomaly descent, they arise from the anomaly
polynomial as an intermediate step in the computation of the consistent anomaly,
while they provide the simplest anomaly inflow mechanism, canceling the
anomaly on a co-dimension-one boundary or defect. In both cases, the BZ current
emerges from the Chern-Simons action via the so-called anti-derivative
operation, providing the key ingredient for converting the consistent
currents in the diffeomorphism and supersymmetry Ward identities to covariant ones.

The rest of the paper is organized as follows. We start with a preliminary
review of the effective action, anomalies, and the standard anomaly
descent in Section \ref{sec:background}, highlighting the role of the BZ current.
We also revisit the standard anomaly descent procedure and the corresponding
BRST algebra as a warm up to our subsequent analysis. In Section \ref{sec:diff}
we review the diffeomorphism anomaly and contrast it with the Lorentz one.
In particular, we explain why the anomaly descent of the Christoffel connection
computes the former, while that of the spin connection computes the Lorentz
anomaly. However, a discussion of the BRST mechanism that underlies this
distinction is deferred to Section \ref{sec:descent}, where we present
the anomaly descent and the BRST algebra in a more general context that allows
us to accommodate supersymmetry as well.

Section \ref{sec:susy} addresses some general features of the supersymmetry Ward
identity and, in particular, how the anomalies associated with other
symmetries give rise to new terms. The BZ current appears precisely due
to these terms, and covariantizes the relevant consistent currents of
gauge/flavor symmetries. Unlike the diffeomorphism case, however, it
does not stop there. The WZ consistency conditions demand further
contributions with more gauginos that are gauge/flavor-invariant.
An interesting question is what should happen exactly to these
induced anomalous terms if some external inflow mechanism is introduced
to cancel the gauge/flavor anomaly to begin with. Generally, the natural expectation
that inflow must cancel all such terms is confirmed up to the BZ term,
showing that the supersymmetry transformation of the effective action becomes gauge invariant
when combined with the bulk inflow term.

Section \ref{sec:inflow} takes up the remaining question of
Section \ref{sec:susy}, on the gauge/flavor invariant part of the induced
anomalous terms. We show that in the case of a co-dimension-one inflow from
a supersymmetric Chern-Simons action, the cancellation is complete and the
combined action, i.e. the effective action and the bulk supersymmetric
Chern-Simons action is invariant under supersymmetry transformations
as long as the underlying gauge/flavor anomaly is canceled
This also means that one can compute the entire supersymmetrized anomaly
simply from the variation of a Chern-Simons action in one dimension higher,
with appropriate supersymmetry, which naturally becomes a boundary term.
Such a co-dimension-one inflow is not the most general form of inflow,
but when it exists, it presents perhaps the most efficient way to
compute the supersymmetric completion.

Finally, we come back to the anomaly descent, or more precisely the
BRST algebra thereof, in Section \ref{sec:descent}, and extend
the standard BRST algebra to accommodate these phenomena systemically.
Although not widely recognized, this modification is actually necessary
to elevate the WZ consistency conditions for diffeomorphisms to
the BRST algebra. A key observation is that  the content of the BRST gauge field does not need to match precisely the structure of the BRST
operator in order  for a descent mechanism to emerge and provide a solution of the WZ
consistency conditions. One can get a first glimpse of the necessity of this
from the diffeomorphism anomaly descent. The BRST transformation
involves both translations and rotations while the usual descent
involves ghosts for the rotational part, say the local
Lorentz transformation, only.

This extension proves essential also in the context of supersymmetry,
and we use this formulation to determine the general structure of
supersymmetrized anomalies. Up to the very first response of the
anomaly to supersymmetry, which is expressed in terms of the BZ current
and described in detail in Section \ref{sec:susy}, the descent procedure
is universal and independent of spacetime dimension. The additional
gauge/flavor invariant terms are sensitive to the precise supersymmetry
multiplet and spacetime dimension and are left implicit. We end Section \ref{sec:descent} by demonstrating that for certain multiplets the generalized descent
determines the supersymmetric Chern-Simons action responsible for the inflow mechanism in Section \ref{sec:inflow}. We have tried to keep the discussion  in Section  \ref{sec:descent} general and minimal, but for completeness we have collected some useful related notions that have appeared in the past literature in the Appendix \ref{sec:gendescent}.

\section{An Overview of Anomalies}
\label{sec:background}

While gauge anomalies and anomalies of global symmetries come with
very different consequences, the classification and computation
of anomalies does not really distinguish between the two classes. Since
these anomalies arise from the path integral of chiral fields, be they
fermions or self-dual tensors, rather than of the gauge fields,
we may as well compute these anomalies on an equal footing by considering
the relevant gauge fields to be all external and the symmetry to be
global. Only at the end of computation, do we worry about the
cancellation of the anomaly if the relevant symmetry is gauged.
Much of what we review here can be traced back to Ref.~\cite{Bardeen:1984pm}.

\subsection{Effective Action and Anomalous Ward Identities}

For a general discussion, let us introduce the effective action
$W(\cA)$ which is a result of path integrals of all chiral fields
coupled to the ``gauge fields,"  collectively denoted as $\cA$,
\bea
e^{-W(\cA)}=\int[\cd\Psi]\;e^{-S(\Psi;\cA)} \ ,
\eea
where the chiral fields responsible for the anomaly are denoted collectively by $\Psi$. Suppose that the action $S(\Psi;\cA=0)$
is invariant under a global symmetry $\delta$ acting on $\Psi$.

This can be then formally elevated to a local symmetry of
$S(\Psi;\cA)$ with the ``gauge field" $\cA$ introduced, so that
\bea
S(\Psi+\delta\Psi;\cA+\delta \cA) = S(\Psi;\cA)\ ,
\eea
or equivalently
\bea
S(\Psi+\delta\Psi;\cA) - S(\Psi;\cA)  = S(\Psi;\cA-\delta\cA)- S(\Psi;\cA) = -\delta \cA\cdot \cJ
\eea
at the linear order, with the sign convention
\bea
\frac{\delta S}{\delta \cA} =  \cJ\,.
\eea
If the path integral measure is invariant under such
transformations as well, we have
\bea
\int[\cd\Psi]\;e^{-S(\Psi;\cA)}=\int[\cd(\Psi+\delta\Psi)]\;e^{-S(\Psi+\delta\Psi;\cA+\delta\cA)} =\int[\cd\Psi]\;e^{-S(\Psi;\cA+\delta\cA)}\ ,
\eea
where, for the second equality, we used that $\Psi$ is a dummy
variable for the integral. However, the anomaly arises precisely because the path integral measure is not invariant,
so the very first step fails.

A gauge transformation $\delta_\Phi $, say,
$\delta_\Phi\cA = \rmd \Phi+\cdots$, induces
\bea
\delta_\Phi W(\cA) =\delta_\Phi \cA\cdot \frac{\delta W}{\delta \cA} = -\Phi \cdot\langle\nabla_\mu\cJ^\mu\rangle \ ,
\eea
whose anomalous value is supposed to be
captured by the so-called consistent anomaly,
\bea
\Phi \cdot\langle\nabla_\mu\cJ^\mu\rangle= \int {\bf w}_{d}^{(1)}(\Phi; \cA)\ ,
\eea
where  ${\bf w}_{d}^{(1)}$ is a local functional obtained from anomaly descent. Throughout this paper we denote these standard anomalies by $G(\Phi;\cA)$. We will give a brief review of
the relevant manipulations in the following subsection.

More generally,  however, external gauge fields
of one symmetry might interfere with such a Ward identity for some
other symmetries. The simplest example of this is diffeomorphisms
$\delta_\xi$;  $\delta_\xi W$ can be expressed
generally as
\bea
\delta_\xi W =\delta_\xi \Gamma \cdot\frac{\delta W}{\delta \Gamma} +\sum_{\cA'\neq\Gamma} \delta_\xi \cA' \cdot\frac{\delta W}{\delta \cA'}\ ,
\eea
where the sum is over the other external gauge fields. $\Gamma$ is the Christoffel connection; $\delta_\xi\Gamma$ is qualitatively different from ordinary gauge transformations since it involves translational components as well. See Section~\ref{subsec:DA}.

In terms of the path integral we have
\bea
\delta_\xi e^{-W} = \int[\cd\Psi]\; \left(\xi_\nu\cdot \nabla_\mu T^{\mu\nu}(\Psi) - \sum'\delta_\xi \cA'\cdot \cJ'(\Psi)\right)
e^{-S(\Psi;\Gamma,\cA')}
\eea
with the energy-momentum tensor $T_{\mu\nu}$.
What is the path integral interpretation of this expression? Note that $\xi_\nu \cdot \nabla_\mu T^{\mu\nu}(\Psi)$ downstairs is constructed
out of the dynamical fields and acts on all things made up of $\Psi$.
Since all currents are  vectorial, the operator $\nabla_\mu T^{\mu\nu}$
acts on $\cJ'$ of the $\cA'\cdot \cJ'$
term as well, which means that $\xi_\nu\cdot \nabla_\mu T^{\mu\nu}(\Psi)$ by itself
will not leave the effective action invariant, even in the absence of an anomaly, but a counteracting rotation of $\cA$ must accompany the transformation. The second term does exactly this, so that $-\delta_\xi W$ is the quantity that is captured by the conventional anomaly, leading to the  Ward identity
\bea\label{diffeo-Ward}
\xi_\nu\cdot\langle \nabla_\mu T^{\mu\nu}\rangle - \sum'\delta_\xi\cA' \cdot \langle  \cJ'\rangle=G_{\rm diffeo}(\xi;\cA)\,.
\eea
Later we will also review the diffeomorphism anomaly $G_{\rm diffeo}(\xi;\cA)$.
As such, it is important to keep track of the difference between
the anomaly, $\delta W$, and the divergence of the would-be-conserved current,
$\langle \nabla_\mu \cJ^\mu\rangle$.

\subsection{Anomaly Descent and Bardeen-Zumino Currents}

The anomaly descent solves the following functional
equation for a local functional $G(\Phi;\cA)$
\bea
\delta_{\Phi_1} \delta_{\Phi_2}W  - \delta_{\Phi_2} \delta_{\Phi_1}W  =  \delta_{[\Phi_1,\Phi_2]}W \ ,
\eea
where 
$\delta_{\Phi}$ is a gauge transformation of the
relevant connection $\cA$. This comes with the sign choice,
\bea
\delta_\Phi \cA = \rmd \Phi+[\cA,\Phi]\ ,\qquad \delta_\Phi \cF = [\cF,\Phi]\ ,
\eea
and the convention that $\delta$ does not act on the
gauge parameters, so that $\delta_{\Phi_1}\delta_{\Phi_2} \cF=[[\cF,\Phi_1],\Phi_2]$.
One could have various different conventions such as $\delta$
acts on parameters, which could be more natural depending
precisely on the computation at hand.

In the BRST formulation \cite{Becchi:1975nq},
we replace $\Phi$ by a Grassmann odd gauge function $v$ and recast
the gauge transformation into the anticommuting version $\bs$ which acts as
\bea\label{BRST}
\bs \cA &= &-{\rmd} v-\cA v-v \cA\ ,\cr
\bs \cF &=& \cF v- v\cF \ ,\cr
\bs v &= &-v^2\,.
\eea
We treat ${\tx d} $ and $\bs$ on an equal footing, and the same with $\cA$ and $v$,
and assign BRST odd grading on all of these. One can recover, say, from $\bs\cA$,
the usual gauge transformation $\delta_\Phi \cA={\tx d} \Phi+\cdots$ if we write
$v=\chi\Phi$ and pull the Grassmann odd constant $\chi$ all the way
to the left. The same happens with $\bs\cF$ since $\cF$ is a two-form and thus even.

The standard anomaly descent is based on these nilpotent
operators $\tx d$, $\bs$, and $\tx d +\bs$, which obey
\be\label{nilpotentold}
\tx d^2=0,\qquad \bs^2=0,\qquad (\tx d+\bs)^2=0 .
\ee
Recall that the nilpotency of $\tx d$ implies that
the field strength $\cf\equiv \tx d\ca+\ca^2$ satisfies
the Bianchi identity $\tx d \cf+\ca\cf-\cf\ca=0$. The two
key observations were the ``Russian formula" of \cite{Manes:1985df}
\be\label{Russian}
\Hat\cf\equiv (\tx d+{\bf s})\Hat\ca+\Hat\ca^2=\cf,\qquad \Hat\ca\equiv\ca+v,
\ee
and the generalized Bianchi identity
\be
(\tx d+\bs)\Hat\cf+\Hat\ca\Hat\cf-\Hat\cf\Hat\ca=0\ .
\ee
The final ingredient is an anomaly polynomial
$P_{d+2}$ such that
\bea
0= {\tx d}  P_{d+2}(\cF)\ ,
\eea
and the related Chern-Simons form ${\bf w}_{d+1}(\cA,\cF)$ that satisfies $P_{d+2}(\cF)={\tx d} {\bf w}_{d+1}(\cA,\cF)$. Similarly, the same polynomial in $\Hat \cF$, $P_{d+2}({\Hat \cF})$, is $(\tx d+\bs)$-closed.

One  generates a solution to the WZ consistency condition by considering
an expansion in $v$ of the right hand side of the identity
\bea\label{descent}
P_{d+2}(\cF) &=&P_{d+2}(\Hat \cF) \cr\cr
&=&({\tx d} +\bs) {\bf w}_{d+1}(\cA+v,\cF) \cr\cr
&=&P_{{\tx d} +2}(\cF) + \bs{\bf w}_{d+1}^{(0)} (\cA,\cF) +  {\tx d} {\bf w}_d^{(1)}(v;\cA,\cF)+O(v^2)\ ,
\eea
where the numeral superscript  keeps track of the power of $v$,
\bea
 {\bf w}_{d+1}(\cA+v,\cF) = \sum_{k\ge 0 } {\bf w}_{d+1-k}^{(k)}(v;\cA,\cF)\,.
\eea
As noted above, the translation back to the bosonic version requires
$v=\chi\Phi$ and moving the Grassmann odd piece $\chi$ all the way
to the left in all expressions. For example,
\bea
\chi \,\delta_\Phi {\bf w}_{d+1}(\cA,\cF) = \bs {\bf w}_{d+1}^{(0)}(\cA,\cF)\biggr\vert_{v\;\rightarrow\; \chi \Phi}
= -{\tx d}  {\bf w}_{d}^{(1)}(\chi \Phi;\cA,\cF) =\chi\, {\tx d}  {\bf w}_{d}^{(1)}(\Phi;\cA,\cF)\,.
\eea

At order $v^2$ the equality in (\ref{descent}) yields
\bea
\bs\int {\bf w}_{d}^{(1)} (v;\cA,\cF) = -\int {\tx d} \left({\bf w}_{d-1}^{(2)}(v;\cA,\cF)\right) =0\,,
\eea
which implies that the integral of ${\bf w}_d^{(1)}$ gives the desired solution to the
consistency condition.

To see this it suffices to take $v=\chi_1\Phi_1+\chi_2\Phi_2$
with Grassmann odd, $\chi_{1}\neq \chi_2$. $\delta_\Phi$ acts only
on fields, while $\bs$ is designed to act on $v$ as well, such that the kernel
of $\bs$ solves the WZ consistency condition, modulo exact
terms. This brings us to the usual anomaly descent solving the consistency
condition,
\bea
-\delta_\Phi W= G(\Phi;\cA)\equiv \int {\bf w}_d^{(1)}(\Phi;\cA,\cF) \ ,
\eea
with
\bea
\delta_\Phi {\bf w}_{d+1}(\cA,\cF) = {\tx d}  {\bf w}_{d}^{(1)}(\Phi;\cA,\cF)
\eea
via an entirely bosonic descent procedure.

A well-known ambiguity in this procedure occurs in the very first step,
\bea
P_{d+2}(\cF) &=&{\tx d}  {\bf w}_{d+1}(\cA,\cF)\ ,
\eea
when a mixed term of type $ P_{2n}(\cF_1)\wedge P_{d+2-2n}(\cF_2) $
is present in the anomaly polynomial. The right hand side has the
ambiguity
\bea
\cdots + \alpha {\bf w}_{2n-1}(\cA_1,\cF_1)\wedge P_{d+2-2n}(\cF_2)+
\beta P_{2n}(\cF_1)\wedge  {\bf w}_{d+1-2n}(\cA_2,\cF_2)+\cdots
\eea
with $\alpha+\beta=0$, so that
\bea
{\bf w}_{d}^{(1)}(\Phi_1;\cA,\cF)&=&\cdots+ \alpha {\bf w}_{2n-2}^{(1)}(\Phi_1, \cA_1,\cF_1)\wedge P_{d+2-2n}(\cF_2)+\cdots \ , \cr\cr
{\bf w}_{d}^{(1)}(\Phi_2;\cA,\cF)&=&\cdots+ \beta P_{2n}(\cF_1)\wedge  {\bf w}_{d-2n}^{(1)}(\Phi_2;\cA_2,\cF_2)+\cdots\,.
\eea
The ambiguity is a matter of a local counter term in $d$ dimensions,
so one can choose which in a more convenient description. What must
be noted here, for later purposes, is that even with such mixed
terms and this ambiguity thereof, we have the invariance,
\bea\label{1v2}
\delta_2 \int {\bf w}_{2n-2}^{(1)}(\Phi_1;\cA,\cF)&=&0
\eea
and vice versa, between a pair of symmetries that are mutually
commuting; $\cA_2$ does not appear explicitly  in
${\bf w}_{2n-2}^{(1)}(\Phi_1;\cA,\cF)$. Only $\cF_2$ enters
and only inside an appropriate trace.

Here we wish to explore a bit more how the anomaly
descent reacts to arbitrary shifts of $\cA\rightarrow \cA+ a$.
It is useful to introduce a nilpotent BRST even operator $\Delta_a$, which acts on any of the above functions of $\cA$, $\cF$, and $v$ as
\bea\label{Delta}
\Delta_a = {\tx d} \, l_a +l_a {\tx d}
\eea
with the replacement ${\tx d} \cA \rightarrow \cF-\cA^2$ understood. The so-called anti-derivative $l_a$ acts as
\bea
l_a\cA =0\ ,\qquad l_a\cF=a\ , \qquad l_av=0 .
\eea
We must treat $l_a$ as a BRST-odd operator naturally,
since $a$ is BRST-odd just like $\cA$.
On the other hand, the anti-derivative $l_a$ also obeys
\bea
l_a \bs +\bs\, l_a=0\ ,
\eea
for which we also need to remember that $a$, as a difference
between two connections $\cA$ and $\cA+a$, is assumed to
transform covariantly, $\bs a=-av-va$.

The action of $\Delta_a$ on the following two quantities is of
some interest,
\bea
\Delta_a\left( {\bf w}_{d+1}^{(0)}(\cA,\cF)\right)&=& {\tx d} \left(l_a \,[{\bf w}_{d+1}^{(0)}(\cA,\cF)]\right) +l_a\left(P_{d+2}(\cF)\right)\ ,\cr\cr
\Delta_a\left( {\bf w}_{d}^{(1)}(v;\cA,\cF)\right)& =& {\tx d} \left( l_a\,[{\bf w}_d^{(1)}(v;\cA,\cF)]\right)  + l_a \left(-\bs\, [{\bf w}_{d+1}^{(0)}(\cA,\cF))]\right)\,.
\eea
Recall that, with
\bea
a\cdot \cX \equiv \int l_a \,[{\bf w}_{d+1}^{(0)}(\cA,\cF)]\ ,
\eea
$\cX$ is the so-called Bardeen-Zumino current which can be added to $\cJ$
to turn the latter consistent current into its covariant version.

From this, we find
\bea
\Delta_a \int {\bf w}_{d}^{(1)}(v;\cA,\cF) =  \bs\int l_a({\bf w}_{d+1}^{(0)})
\eea
with a vanishing boundary condition at asymptotic infinity
of the spacetime. Again, coming back to the original bosonic form,
\bea
\Delta_a \int {\bf w}_{d}^{(1)}(\Phi;\cA,\cF)
=\delta_\Phi\int l_a\left({\bf w}_{d+1}^{(0)}(\cA,\cF)\right)\,.
\eea
The integrand on the right hand side is
proportional to an inner product of $a$ with the
so-called BZ current,  $\cX$, as
\bea
\Delta_a \int {\bf w}_{d}^{(1)}(\Phi;\cA,\cF) = \delta_\Phi ( a\cdot \cX  )\ .
\eea
Recall that this BZ current shifts the
consistent current $\cJ$ additively into the covariant
current $\cJ_{\rm cov} = \cJ + \cX$ \cite{Bardeen:1984pm}.

\section{Diffeomorphism Anomaly and Covariant Currents}
\label{sec:diff}

Let us now be more specific and split the gauge transformation
into two classes; those associated with spacetime transformations,
such as diffeomorphisms and supersymmetry, and the internal
gauge/flavor symmetries. The notation we will adopt is
\bea
\cA =(\Gamma,\psi ; A)\ ,\qquad \Phi = (\xi,\epsilon ; \vartheta) \ , \qquad \cJ=(T, S ;J)\ ,
\eea
where the external gravitino $\psi$ and the supersymmetry
current $S$ would be present if the theory is supersymmetric.
For general discussion of the diffeomorphism anomaly and the Ward
identity thereof, we will adopt the Christoffel connection, $\Gamma$,
although, for supersymmetry Ward identities in Section~\ref{sec:inflow},
the spin connection $\o$ is more natural and will be used instead.

We are now quite used to the fact that the variation of the effective
action induces the consistent anomaly which solves the WZ consistency
condition. Despite some confusions in the early literature, the relation
between the consistent and the covariant anomaly was understood
already in 1980's \cite{Bardeen:1984pm}. The consistent
current $\cJ$ transforms non-covariantly, but using the BZ term,
$\cX$, which is composed entirely of the external gauge field, a
covariant current can be found, namely $\cJ_{\rm cov}=\cJ+\cX$. The divergence
of this combination is called the covariant anomaly.
In turn, the covariant anomaly proved to be much more practical for
computational purposes, eventually yielding the well-known anomaly
polynomials \cite{AlvarezGaume:1983ig}. The physical role of the
 covariant current has been much less prominent. The purpose of
 this section is to show that the covariant current must, in fact,
 appear in the Ward identity setting generically.

The message is actually very simple. The action of diffeomorphisms on the
effective action should induce the consistent current of the gauge/flavor symmetry
as well, since the former acts universally on all operators. On the other hand,
the diffeomorphism Ward identity itself must be invariant
under such internal symmetries. As we shall see, the shift to $J_{\rm cov}=J+X$
occurs naturally. Another such symmetry transformation that acts  on all
currents universally is supersymmetry, which is the main subject of this
note. This section will provide a simpler illustration for the latter,
more involved case, to which we will turn in Section \ref{sec:susy}.

\subsection{Diffeomorphism Anomaly }
\label{subsec:DA}

As a simple illustration, let us start with a theory with
a gauge symmetry $\delta_\vartheta$ and consider how the diffeomorphism
Ward identity is affected by the anomalous gauge sector. For this, let us
start with
\bea
\cA =(\Gamma; A)\ ,\qquad \Phi = (\xi; \vartheta) \ , \qquad \cJ=(T;J) \ ,
\eea
where the Christoffel connection can be conveniently elevated to
a connection 1-form \cite{Bardeen:1984pm},
\bea
(\Gamma)^{\;\;\alpha}_{\beta} \equiv -\Gamma^\alpha_{\mu\beta}\, dx^\mu
\eea
with the curvature 2-form
\bea
R^{\;\;\alpha}_{\beta} = ({\tx d} \Gamma +\Gamma\wedge \Gamma)^{\;\;\alpha}_{\beta}\ .
\eea
This translates to the usual Riemann curvature as
\bea
\cR^{\alpha}_{\;\;\beta\mu\nu} = (R_\beta^{\;\;\alpha})_{\nu\mu}\,.
\eea
The two are clearly the same component-wise once the symmetry
properties of the Riemann tensor are invoked. Diffeomorphisms act
on the Christoffel connection as
\bea
\delta_\xi \Gamma &=&\cL'_\xi\Gamma +{\tx d}_\Gamma (-\partial \xi)\ ,
\eea
where  $\cL'_\xi$ is meant to be aware only of the 1-form index
of $\Gamma$. Note how the  transformation is split into two pieces
and how the second piece can be regarded as a $GL(d)$ gauge transformation with
$\Phi_\mu^{\;\;\nu} = -\partial_\mu \xi^\nu$ of $\Gamma$.\footnote{
Its counterpart for the spin connection $\omega$ is
\bea
\delta_\xi \omega &=&\cL_\xi''\omega +{\tx d}_\omega\hat\xi_K,
\eea
where $\cL_\xi''$ is again aware only of the 1-form index
of $\omega$.  The $SO$ matrix $ \hat\xi_K^{ab} \equiv - \nabla^{[a}\xi^{b]}
-\xi^\mu\Omega_\mu^{ab}$ is known as the Kosman lift \cite{Kosman}. It has been established
that the diffeomorphism anomaly is, modulo a counterterm, equivalent
to the purely rotational $SO(d)$ Lorentz anomaly \cite{Bardeen:1984pm,Bertlmann:1996xk}.
This is why we can usually get away with relying on and treating $\omega$ as
if it is one of the internal gauge fields. However, the Ward identity in
question is that of diffeomorphisms, for which $\Gamma$ proved to be
more suitable. }

Another difference compared to gauge transformations is how the
commutator of $\delta_\xi$ works. On any covariant tensor, $V$,
we have
\bea
[\delta_\zeta,\delta_\xi]V = \cL_\xi(\cL_\zeta V) - \cL_\zeta(\cL_\xi V) =\cL_{[\xi,\zeta]}V= -\delta_{[\zeta,\xi]}V\ ,
\eea
where, as with the previous gauge variations, we let $\delta$ act
on field variables only and not on parameters.
This leads to the gravitational WZ consistency condition,
\bea\label{ddWZ}
\delta_\zeta \delta_\xi W  -\delta_\xi \delta_\zeta W
=  \delta_{-[\zeta,\xi]}W\ ,
\eea
where one must note the unusual sign on the right hand side.
Bardeen and Zumino addressed how to solve this in their seminal paper \cite{Bardeen:1984pm}
and found that the usual anomaly descent procedure solves this constraint,
\bea
-\delta_\xi W = G_{\rm diffeo}(\xi;\Gamma,R;F) \equiv \int {\bf w}_d^{(1)}(-\partial\xi; \Gamma, R;F)\,,
\eea
where we treat $\Gamma$ as a $GL(d)$ connection and
perform the descent as if this $GL(d)$ is an internal gauge symmetry.

This can be seen from
\bea
\delta_\zeta G_{\rm diffeo}(\xi;\Gamma,R;F) 
= \int \left(\hat \cL'_\zeta +\delta_{-\partial\zeta}^{GL(d)}\right) {\bf w}_d^{(1)}(-\partial\xi; \Gamma, R;F)\,,
\eea
where the hat and the prime on $\hat \cL_\zeta'$ mean that it acts only on
fields and responds only to the differential form indices in $\Gamma$, $R$,
and $F$. In particular, it perceives the entire ${\bf w}_d^{(1)}$ as
a differential $d$-form.
The second variation is a purely rotational $GL(d)$ gauge rotation,
so one can invoke a $GL(d)$ WZ consistency condition also
obeyed by $G_{\rm diffeo}(\xi;\Gamma,R;F)$,
\bea
&& \delta_{-\partial\zeta}^{GL(d)} \int {\bf w}_d^{(1)}(-\partial\xi; \Gamma, R;F)
-\delta_{-\partial\xi}^{GL(d)} \int {\bf w}_d^{(1)}(-\partial\xi; \Gamma, R;F)\cr\cr
&=& \int {\bf w}_d^{(1)}([-\partial\zeta, -\partial\xi]; \Gamma, R;F)\,.
\eea
On the other hand, the first derivative operator $\hat \cL'_\zeta$ can be made
to act, by integration by parts and with the induced sign flip, on the
$GL(d)$ parameter,  $-\partial\xi$,  as if it is a (matrix-valued) scalar function.

Combining the two, we find
\bea
&&\delta_\zeta G_{\rm diffeo}(\xi;\Gamma,R;F) -\delta_\xi G_{\rm diffeo}(\zeta;\Gamma,R;F) \cr\cr
&= &\int {\bf w}_d^{(1)}(\partial(\zeta^\mu \partial_\mu\xi - \xi^\mu \partial_\mu\zeta); \Gamma, R;F)\ ,
\eea
from which
\bea
\delta_\zeta G_{\rm diffeo}(\xi;\Gamma,R;F)  -\delta_\xi G_{\rm diffeo}(\zeta;\Gamma,R;F)
=  G_{\rm diffeo}(-[\zeta,\xi];\Gamma,R;F)
\eea
follows immediately, so that $- \delta_\xi W =  G_{\rm diffeo}(\xi;\Gamma,R;F)$
solves (\ref{ddWZ}) with the obvious matrix multiplication rule. Note that in
the above $GL(d)$ anomaly descent, $G_{\rm diffeo}$ depends on $F$
but not on $A$, and only via symmetric traces of $F$, which implies that
\bea
\delta_\vartheta G_{\rm diffeo}(-\partial\xi;\Gamma, R;F)=0\,.
\eea
In Section \ref{sec:descent}, we will see how this diffeomorphism
anomaly can be naturally embedded into a BRST algebra, but with a
twist, which will lead us to a modification that is needed for
the anomaly descent in the presence of supersymmetry as well.

\subsection{Covariant Currents in the Diffeomorphism Ward Identity}

What about diffeomorphisms acting on the gauge anomaly? The entire
anomaly is a local $d$-form integral which is linear in $\vartheta$
and polynomial in $A$, $F$, and $R$, all of which are tensors with respect
to general coordinate transformations. Given that $\delta_\xi$ acts on
$A,F,R$ as $\cL_\xi$ but does not act on $\vartheta$, we may integrate by parts to obtain
\bea
\delta_\xi \int {\bf w}_{d}^{(1)}(\vartheta;A,F;R) = \int {\bf w}_{d}^{(1)}(-\cL_\xi\vartheta;A,F;R) = - G(\cL_\xi \vartheta;A,F;R)\,.
\eea
As such, the anomaly descents that we have accumulated so far obey
\bea
\delta_\vartheta G_{\rm diffeo}(-\partial\xi;\Gamma,R;F)-\delta_\xi G(\vartheta;A,F;R) = G(\cL_\xi \vartheta;A,F;R)
\eea
and thus solve a consistency condition
\bea
\delta_\vartheta \delta_\xi W - \delta_\xi \delta_\vartheta W = \delta_{\cL_\xi \vartheta} W\,.
\eea
This last expression can be seen to be the correct WZ consistency condition as
\bea
(\delta_\vartheta\delta_\xi-\delta_\vartheta\delta_\xi) W
&=&  \left(\delta_\vartheta(\xi\lrcorner {\tx d} A+{\tx d}(\xi\lrcorner A))
- \delta_\xi({\tx d}_A \vartheta)\right)\cdot \frac{\delta W}{\delta A} \cr\cr
&=&  \left(  ({\tx d} \cL_\xi \vartheta)    +\cL_\xi ([A,\vartheta])
- [\cL_\xi A ,\vartheta]\right)\cdot \frac{\delta W}{\delta A} \cr\cr
&=&  \delta_{\cL_\xi \vartheta}A\cdot \frac{\delta W}{\delta A}
\:=\: \delta_{\cL_\xi \vartheta} W
\eea
if we  keep in mind that $\delta$'s do not act on parameters.

Here, we wish to address the transformation property of
$\delta_\xi W$
\bea\label{diffeo}
-\delta_\xi W =\xi_\nu\cdot\langle \nabla_\mu T^{\mu\nu}\rangle - \sum \delta_\xi A \cdot\langle J\rangle
\eea
under $\delta_\vartheta$. We have seen earlier that the mixed WZ consistency condition
\bea
\delta_\vartheta (\delta_\xi W) -\delta_\xi(\delta_\vartheta W) = \delta_{\cL_\xi \vartheta} W
\eea
constrains $\delta_\xi W$, and furthermore, given the specific
form of the gauge anomaly descent,
\bea\label{invariance}
\delta_\vartheta (\delta_\xi W) =0 \ ,  \qquad \delta_\xi(\delta_\vartheta W) + \delta_{\cL_\xi \vartheta} W =0
\eea
have to hold separately. We have seen the former above while the latter
can be seen more explicitly as follows. Since $\delta_\xi$ is supposed
to act only on field variables and not on the transformation parameters,
such as $\vartheta$, we have
\bea
\delta_\xi(\delta_\vartheta W) + \delta_{\cL_\xi \vartheta} W &=&- \delta_\xi\int {\bf w}_d^{(1)}(\vartheta;A;F;R)-
\int {\bf w}_d^{(1)}(\cL_\xi\vartheta;A;F;R)\cr\cr
&=& -\int \cL_\xi {\bf w}_d^{(1)}(\vartheta;A;F;R)\cr\cr
&=& -\int {\tx d} (\xi\lrcorner {\bf w}_d^{(1)})+\xi\lrcorner {\tx d} {\bf w}_d^{(1)}\;=\;0\ ,
\eea
where for the last equality we have used ${\tx d} {\bf w}_d^{(1)}=0$ as well,
which appears to contradict the formal procedure we used for the
descent, but does hold once we come back to physical quantities
and coordinates relevant to a $d$-dimensional spacetime; one simply
cannot have a nonzero $(d+1)$-form, i.e., ${\tx d}{\bf w}_d^{(1)}$,
in a $d$-dimensional spacetime.\footnote{This obvious observation
will also be quite useful when we introduce a generalized anomaly
descent procedure for diffeomorphisms and  supersymmetry in Section \ref{sec:descent}.}

How does this reconcile with the appearance of the
consistent gauge current in the Ward identity
from $\delta_\xi W$, as opposed to the covariant one,
in Eq.~(\ref{diffeo})? This is relatively easy to see for
a single Abelian $A$. Note that $\cL_\xi$ acts on $A$
as a vector
\bea
\cL_\xi A = {\tx d} \left(\xi \lrcorner \,A\right) + \xi \lrcorner\, {\tx d} A\ ,
\eea
so, with $F={\tx d} A$,
\bea
-\delta_\xi W =\xi_\nu\cdot\langle \nabla_\mu T^{\mu\nu}\rangle- (\xi \lrcorner\, F ) \cdot\langle J\rangle
+ (\xi \lrcorner \, A) \cdot \langle \nabla_\mu J^\mu \rangle\,.
\eea
On the other hand the last piece is nothing but a pure gauge
anomaly with $\vartheta=-(\xi \lrcorner \, A)$, so it is a local
functional involving $\xi$, $A$, and $F$. This last formula
actually holds for non-Abelian $A$ as well, via $F=dA+A^2$.

After some tedious algebra, using the explicit form of
the gauge anomaly, one realizes that this last term
can be rewritten via the BZ current $X$ and  the curvature
$F$  such that \cite{Jensen:2012kj,Papadimitriou:2017kzw}
\bea
-\delta_\xi W =\xi_\nu\cdot\langle \nabla_\mu T^{\mu\nu}\rangle - (\xi \lrcorner\, F ) \cdot\langle J +X \rangle
=\xi_\nu\cdot\langle \nabla_\mu T^{\mu\nu}\rangle - (\xi \lrcorner\, F ) \cdot\langle J_{\rm cov}\rangle\,.
\eea
As such, the diffeomorphism Ward identity is
manifestly gauge-invariant under $\delta_\vartheta$
\bea
\xi_\nu\cdot\langle \nabla_\mu T^{\mu\nu}\rangle - (\xi \lrcorner\, F ) \cdot\langle J_{\rm cov}\rangle= \int {\bf w}_d^{(1)}(-\partial\xi ; \Gamma, R;F)\,.
\eea
The right hand side is invariant under $\delta_\vartheta$ thanks
to the general fact in \eqref{1v2}, so the above covariantization
of the gauge current is inevitable. Although the necessary
conversion was performed explicitly for Abelian $A$, $\delta_\vartheta(\delta_\xi W)=0$
means that  the same conversion of 
the anomaly term to the $(\xi\lrcorner F)\cdot X$
occurs for non-Abelian $A$'s as well. The latter fact will
appear again crucially when we discuss the supersymmetrized
anomaly descent in Section~\ref{sec:descent}.

\section{Supersymmetry and Bardeen-Zumino currents}
\label{sec:susy}

Let us now turn to the supersymmetry Ward identity and discuss the
local fermionic terms induced by anomalies in other symmetries, which
are therefore controlled by the same anomaly coefficient. Such terms
induced by the gauge/flavor anomaly have been discussed long ago in \cite{Piguet:1980fa,Itoyama:1985qi,Guadagnini:1985ea,Itoyama:1985ni,Girardi:1985hf,Hwang:1985tm,Altevogt:1987qe,Kaiser:1988zg,Altevogt:1987fx,Altevogt:1989fw,Baulieu:2008id,Bzowski:2020tue}, while analogous terms induced by the gravitational anomaly in two dimensions were first identified in \cite{Tanii:1985wy,Howe:1985uy}, and locally supersymmetric contributions to gauge/favor or $R$-symmetry anomalies were discussed
more recently in \cite{Papadimitriou:2017kzw,An:2017ihs,Papadimitriou:2019gel,
Papadimitriou:2019yug,An:2019zok,Kuzenko:2019vvi,Katsianis:2019hhg,Katsianis:2020hzd,Nakagawa:2021wqh}.

It was understood early on that if a superspace
description of the external gauge multiplet exists, such local fermionic
terms can be understood as a consequence of the WZ gauge. In particular,
if one keeps all auxiliary fields and maintains gauge symmetry at a fully
supersymmetric level, one could move these terms to the left hand side
of the Ward identity. However, there are several contexts where a superspace
or fully off-shell multiplet does not exist, and therefore we do not assume
its existence in the present analysis.

As we will see, the local fermionic terms in the supersymmetry Ward identity
induced by anomalies of other symmetries can be separated into two distinct
types: gauge invariant ones and non-invariant ones \cite{Itoyama:1985qi}.
In the analogous case of the diffeomorphism Ward identity, the latter was
a simple consequence of the supersymmetry transformation of the external
gauge fields and is completely determined by the BZ current. In the
supersymmetry Ward identity, this non-invariant piece is again expressed via
the BZ current but the way it emerges in the WZ gauge is rather different;
it appears as an additional inhomogeneous contribution to the Ward identity,
as demanded  by the consistency condition that involves both gauge/flavor
transformations and supersymmetry transformation.

The other gauge-invariant pieces, an analogue of which does not exist in the
diffeomorphism Ward identity, depends more sensitively on the supersymmetry
algebra. What remains true, however, these are connected to the above non-invariant
contribution again via the consistency condition. In this section, we discuss
the general shape of these local fermion terms, with emphasis on
the non-invariant BZ current contributions, and how exactly the same structure
appears from the inflow mechanism, regardless of the precise details of the
latter.

Since the non-invariant term involving the BZ current is entirely determined
by the gauge/flavor anomaly, to begin with, one expects that this induced BZ
term and the subsequent invariant pieces would be also canceled eventually
if an anomaly inflow \cite{Callan:1984sa} cancels the gauge/flavor anomaly
in question. In principle, therefore, the anomaly inflow may be viewed as
a tool for determining all such fermionic terms, non-invariant and invariant,
if the supersymmetry completion of the bulk action responsible for the inflow
is known.

In general the latter is a tall order, however, given how the bulk action
is typically of higher derivatives. This is true of the simplest inflow mechanism
we use in this section for illustration which is itself somewhat limited, modeled
after the simplest of the M5 brane inflow \cite{Duff:1995wd,Witten:1995em}.
More generally, even more elaborate inflow mechanisms
\cite{Green:1996dd,Cheung:1997az, Minasian:1997mm,Freed:1998tg,Kim:2012wc}
are often unavoidable. Supersymmetry completion of such general topological
terms in string theories or in M-theory are hardly known.

On the other hand, if the inflow originates from a Chern-Simons action in one
dimension higher, the chances are better.
In Section \ref{sec:inflow}, we will delve into several explicit examples of
supersymmetric anomaly inflow of this kind and use them  to determine
the local fermionic terms in the supersymmetry Ward identity induced by anomalies
in other symmetries. The Chern-Simons form ${\bf w}_{d+1}^{(0)}$ determines
the contribution of the gauge/flavor anomaly to the supersymmetry Ward identity
and plays an important role in both supersymmetric anomaly inflow and anomaly
descent. A generalized anomaly descent mechanism that accommodates supersymmetry
will be discussed in Section \ref{sec:descent}.

\subsection{Supersymmetry Ward Identity Revisited}

The external vector multiplet is specified by
\bea
\cA =(\psi ; A)\ ,\qquad \Phi = (\epsilon ; \vartheta) \ , \qquad \cJ=( S ;J)\ ,
\eea
with the gaugino superpartner $\lambda$ of $A$,
\bea
\delta_\epsilon W = \delta_\epsilon A \cdot\frac{\delta W}{\delta A}
+ \delta_\epsilon \lambda \cdot\frac{\delta W}{\delta \lambda}\,.
\eea
As in the case of diffeomorphisms above, we wish to explore what
consequences follow from the WZ consistency condition,
\bea\label{WZSWZ}
\delta_\vartheta (\delta_\epsilon W) - \delta_\epsilon(\delta_\vartheta W) = 0\,.
\eea
The second term on the left is
\bea
-\delta_\epsilon(\delta_\vartheta W) = \delta_\epsilon \int {\bf w}_{d}^{(1)}(\vartheta;A,F)\,.
\eea
Recalling the algebra that leads to the BZ current $X$,
\bea
\Delta_{\delta_\epsilon A}\int {\bf w}_{d}^{(1)}(\vartheta;A,F)  = \delta_\vartheta \int l_{\delta_\epsilon A} \left( {\bf w}_{d+1}^{(0)}(A,F)\right)\ ,
\eea
where $l_{\delta_\epsilon A}$ is the anti-derivative we reviewed earlier,
we realize that
\bea
-\delta_\epsilon(\delta_\vartheta W) = \delta_\vartheta \left(\delta_\epsilon A \cdot X\right)
\eea
again with the BZ current $X$ of the gauge symmetry.
This means that, even in the absence of true anomaly term $G(\epsilon;\psi; A,\lambda)$,
$\delta_\epsilon W $ cannot vanish, but rather must obey
\bea
 -\delta_\epsilon W = \delta_\epsilon A \cdot X +\cdots
\eea
with the ellipses denoting terms that are invariant under $\delta_\vartheta$ and are determined by the consistency condition following from two successive supersymmetry transformations, namely
\be
(\d_\e\d_{\e'}-\d_{\e'}\d_\e)W=(\d_\x +\d_\L)W,\qquad \x^\m\sim \bar\e'\g^\m\e, \quad \L\sim \xi \lrcorner \, A.
\ee
Shortly we will offer an alternate method for this supersymmetry completion
via anomaly inflow, so here we mostly focus on the term involving the BZ current.

An interesting fact is that we can rewrite this Ward identity as \cite{Papadimitriou:2017kzw}
\bea\label{WardSWZ}
\epsilon \cdot \langle \nabla_\mu S^\mu\rangle -
\delta_\epsilon A\cdot \left(\langle J\rangle  +X\right)  - \cdots =0\ ,
\eea
where we have moved the BZ current and its supersymmetry completion, meaning
additional gauge invariant terms whose presence is demanded by the WZ consistency
condition and the very first BZ piece, to the left hand
side. Once this shift of $J$ to $J_{\rm cov}=J+X$ is done,
each component of this Ward identity is individually $\delta_\vartheta$-invariant.
Again we see that $J_{\rm cov}$ appears naturally but the way
this happens here is a little different from the diffeomorphisms
case, where it happened via a rearrangement of the consistent current term.

Here, the main message is that this shift is something to be expected
once we realize that in the Ward identities of spacetime symmetries
the gauge/flavor current appears always in its covariant form. This
also means that the shift is immediate and explicitly known once the
gauge anomaly is known, without having to compute an entirely new set
of diagrams; the shift by the BZ current $X$ is determined
entirely by the $\delta_\vartheta$ anomaly.

The local fermionic terms, $\delta_\epsilon A\cdot X  +\cdots $,
in the supersymmetry Ward identity should be
distinguished from what one might consider an inherent anomaly of
supersymmetry. Nevertheless, this shift of the Ward identity is not
devoid of physical consequences and must be kept track of carefully;
for example, it has known consequences in various supersymmetric
partition function computations that relies on curved spacetime
background and various external fluxes.

In superspace, or in terms of a fully off-shell external vector multiplet,
if such a description exists, the covariant current and its accompanying
supersymmetry partners would have appeared naturally on the left hand side,
on par with the diffeomorphism example of the previous section. This would
allow one to say that the effective action is invariant under supersymmetry
transformations \cite{Itoyama:1985qi}. However, one should be mindful that
such a fully off-shell extension is not always available. Even if one is
available, as would be the case with a smaller supersymmetry content, it is
not clear what one would gain in practice by avoiding the WZ gauge choice.
We might as well concentrate on the nature of these ``anomalous"
contributions to the Ward identity; they would not disappear even in
the superspace version, but merely be attributed differently, and, as has been
seen recently \cite{Papadimitriou:2017kzw,Closset:2019ucb}, can produce a tangible difference in some localization computations.

\subsection{Symmetry Restoration via Gauge/Flavor Anomaly Inflow }

The simplest form of anomaly inflow can be found
from a topological coupling of the type
\bea
S_{\rm inflow} = (-1)^{p}\int_{X_D} H_{p+1}\wedge {\bf w}_{d+1}^{(0)}(\cA)
\eea
in an ambient spacetime $X_D$ with $p=D-d-2$. A $d$-dimensional
worldvolume enters the picture as a magnetic defect to the
spacetime gauge field strength $H_{p+1}={\tx d} A_p+\cdots $ such that
\bea
{\rm d}H_{p+1}=\delta_{M_d}
\eea
and supports an anomalous field content. Suppose that we have an
effective action $W(\cA)$ on $M_d$ whose anomaly polynomial
is $P_{d+2}={\tx d}{\bf w}_{d+1}^{(0)}(\cA)$.

A gauge variation of $S_{\rm inflow}$ leads to
\bea
(-1)^{p}\int_{X_D} H_{p+1}\wedge
\delta_\Phi {\bf w}_{d+1}^{(0)}(\cA)
 = \int_{M_d} {\bf w}_{d}^{(1)}(\Phi;\cA)\ ,
\eea
so that the combined effective action,
\bea
\cW (\cA)= W(\cA) + S_{\rm inflow}
\eea
is gauge-invariant, $\delta_\Phi\cW=0$.
While one must consider more involved versions of this
to cover all known types of anomaly inflow, the essence
of the inflow mechanism is well represented by the example above.

Here, let us ask a slightly different question, namely whether
$\delta_\epsilon\cW$ is gauge invariant. Recall that the
WZ consistency condition demands that 
we have
\bea
-\delta_\epsilon W =\delta_\epsilon A\cdot X +\cdots\ ,
\eea
where the BZ current $X$ can be traced back
to the gauge anomaly $\delta_\vartheta W\neq 0$. If the
latter is canceled by the anomaly inflow from $S_{\rm inflow}$,
it is only natural to expect that $\delta_\epsilon A\cdot X $
on the right hand side of the supersymmetry variation is
also canceled as well, so that the quantity
$\delta_\epsilon\cW$ is gauge invariant.

The relevant quantity to compute is\footnote{We think of the
worldvolume spinors as a chiral projection of spacetime spinors
on $X_D$. Some specific examples of such embedding are discussed in Section \ref{sec:inflow}.}
\bea
\left((-1)^{p}
\int_{X_D} H_{p+1}\wedge \Delta_{\delta_\epsilon A}{\bf w}_{d+1}^{(0)}(\cA)\right)\,.
\eea
Using the descent mechanism we have learned in the previous sections,
this becomes
\bea\label{SVinflow}
&&(-1)^{p}
\int_{X_D} H_{p+1}\wedge ({\tx d} \, l_{\delta_\epsilon A} + l_{\delta_\epsilon A} {\tx d} )\left({\bf w}_{d+1}^{(0)}(\cA)\right) \cr\cr
&=& (-1)^{p}\int_{X_D} H_{p+1}\wedge \left( {\tx d} \left[ \delta_\epsilon A\wedge X\right] +l_{\delta_\epsilon A} P_{d+2}(\cF)\right)\,.
\eea
The first term, when integrated by parts, reduces to
\bea\label{inflowBZ}
 \delta_\epsilon A \cdot X \,.
\eea
Combining the two contributions, we find
\bea
\delta_\epsilon \cW = \delta_\epsilon W + \delta_\epsilon S_{\rm inflow} = 0 +\cdots\ ,
\eea
where the ellipses denote all terms we have neglected so far,
namely the gauge-invariant part of $\delta_\epsilon W$ and the second piece of (\ref{SVinflow}), which does not naturally
reduce to the worldvolume. Therefore, we learn that, modulo
the bulk piece in $\delta_\epsilon S_{\rm inflow}$, the
gauge-invariance of $\delta \cW$ is restored through the gauge
anomaly inflow, such that
\bea
\delta_\vartheta\left( \delta_\epsilon \cW \right) = 0
\eea
on the world-volume.

So, what about the bulk pieces in the variation of $S_{\rm inflow}$?
Once we embed these discussions to supersymmetric theories in the
bulk, we should expect any leftover bulk terms, such as the second piece in
(\ref{SVinflow}) to be canceled by the transformation of the superpartners. As such, in order to complete the study, we
need to start from the fully supersymmetrized form of $\delta_\epsilon W$  and $S_{\rm inflow}$.
For minimal rigid supersymmetry the former is known from e.g. \cite{Itoyama:1985qi,Guadagnini:1985ea,Itoyama:1985ni}
for up to $d=6$, off-shell and in the WZ gauge, while the latter could prove
more involved. After all, such topological terms, $S_{\rm inflow}$,
involve generically higher derivative terms and were often discovered
only via the anomaly cancellation of the entire string theory or M-theory.

On the other hand, there are simpler subclasses where this kind
of question can be asked. One is when the inflow is achieved by a
Green-Schwarz mechanism with $D=d$ and another is where
\bea
S_{\rm inflow}= \int_{X_{d+1}}  {\bf w}_{d+1}^{(0)}(A) +\cdots
\eea
on a bulk $X_{d+1}$ whose boundary is the worldvolume $M_d$.
In fact, the entire class of AdS$_{d+1}$ theories coupled to boundary
CFT belong to such a class. With the proper supersymmetry
instituted in these  $S_{\rm inflow}$, one could ask what
happens to the supersymmetry variation of the combined
action, $\delta_\epsilon \cW$.

We will see in the next section, example by example, that once all
these terms are delineated, we find that not only gauge
invariance is restored
\bea
\delta_\Phi\left(\delta_\epsilon \cW \right)=0
\eea
with $\cW = W+S_{\rm inflow}$, but the combined action is fully
supersymmetric,
\bea
\delta_\epsilon \cW=0.
\eea
Thus, let us call this mechanism of anomaly cancellation a
``supersymmetric anomaly inflow."

\section{Supersymmetric Anomaly Inflow}
\label{sec:inflow}

In the previous section we considered supersymmetric theories with an anomaly,
\bea
-\delta_\vartheta W = \int {\bf w}_d^{(1)}(\Phi;\cA,\cF)\ ,
\eea
for which the supersymmetry Ward identity acquires unexpected
terms on the right hand side, 
\bea\label{susyA2}
-\delta_\epsilon W = \int l_{\delta_\epsilon \cA}{\bf w}_{d+1}^{(0)}(\cA,\cF) +\cdots
\eea
with the usual anomaly descent relating the two right hand sides as
$\delta_\vartheta {\bf w}_{d+1}^{(0)} = {\tx d}  {\bf w}_d^{(1)}$. Although
we have discussed this explicitly for gauge/flavor symmetry here,
a similar structure arises for $R$-symmetry and diffeomorphism
anomalies, which we will revisit in later sections.

On the other hand, given a canceling anomaly inflow term $S_{\rm inflow}$,
we saw that the combined effective action, $\cW=W+S_{\rm inflow}$, satisfies
\bea
\delta_\Phi \cW = 0 \ , \qquad \delta_\Phi(\delta_\epsilon \cW)=0.
\eea
Given that the two right hand sides of $\delta W$ above originate
from a common gauge anomaly, it is quite natural to expect that
\bea
\delta_\epsilon \cW=0\,,
\eea
as long as the inflow action is properly supersymmetrized. In this
section, we demonstrate this mechanism for several examples
where a supersymmetric $S_{\rm inflow}$ is available.

Note that this means that there are alternate ways to compute the
right hand side of (\ref{susyA2}), without having to solve directly the
WZ consistency conditions. In a sense, such an alternate
computation via $S_{\rm inflow}$ can be regarded as a supersymmetrized
anomaly descent especially when $S_{\rm inflow} = S_{\rm CS}$.
This is not to say that one can always achieve the same kind
of universal procedure as in the usual anomaly inflow, since
the crucial supersymmetrization of $S_{\rm CS}$ is often
cumbersome. One complication is that, with a physical inflow
mechanism, given the typically higher dimensional nature
of $S_{\rm inflow}$, both supersymmetry and the field
content thereof tends to be enlarged, beyond what is strictly necessary for $W$.
In the next section we explore a more streamlined and mathematically compact
version of the supersymmetric anomaly descent, inspired by these inflow phenomena. As we show at the end of that section, however, the two approaches coincide for certain multiplets, with the descent procedure determining the supersymmetric Chern-Simons form implementing the co-dimension-one inflow.

In the concrete examples of supersymmetric anomaly inflow that we present in the remaining of this section, we wish to emphasize the utility of such a mechanism as an efficient way to determine the form of supersymmetrized anomalies. To this end, we flip the sign of the Chern-Simons terms relative to the discussion above, $S_{\rm inflow} =- S_{\rm CS}$, in the following such that
\be
\d_\e W=\d_\e S_{\rm CS} \ .
\ee

\subsection{Gauge/Flavor Anomaly Inflow}
\label{sec:flavorAI}

The first example of a supersymmetric anomaly inflow we consider is in the context of flavor anomalies in $d$ dimensions. The anomaly inflow mechanism we focus on relies on the existence of a suitable supersymmetric Chern-Simons action in $d+1$ dimensions. Higher codimension inflow may also be possible in certain cases, such as on the worldvolume of D-branes, but we will not consider this mechanism here.

A supersymmetric Chern-Simons action for a given gauge multiplet in $d+1$ dimensions determines the supersymmetric flavor anomaly in $d$ dimensions for any theory that can consistently couple to the background gauge multiplet obtained from that in $d+1$ dimensions by dimensional reduction. The flavor anomalies for all multiplets with less supersymmetry can be obtained by consistently truncating the resulting $d$ dimensional multiplet.

However, given a theory with $\cn$-extended supersymmetry in $d$ dimensions, there may not exist a corresponding Chern-Simons action with the same amount of supersymmetry in $d+1$ dimensions. In such cases, the $d$ dimensional gauge multiplet must first be embedded in one with a larger amount of supersymmetry, for which a supersymmetric Chern-Simons action does exist. This situation arises, for example, for theories with $(p,q)$ supersymmetry in two dimensions when $p\neq q$. In order to cancel the anomaly through anomaly inflow in such cases, the $(p,q)$ multiplet must be embedded in the closest non-chiral one, namely the flavor multiplet with $(p',p')$  supersymmetry, where $p'=\max(p,q)$. As long as $p'\leq 3$, there exists an $\cn=p'$ supersymmetric Chern-Simons action that provides an anomaly inflow mechanism for the $(p',p')$ anomaly in two dimensions. The original $(p,q)$ anomaly is obtained by multiplet truncation.

\subsubsection{3d to 2d Anomaly Inflow}

The simplest examples of supersymmetric anomaly inflow arise for $d=2$, since the relevant Chern-Simons actions in three dimensions are typically known. The most general 3d Chern-Simons action without matter multiplets involves the $\cn=3$ vector multiplet \cite{Kao:1992ig,Kao:1995gf}, which consists of a gauge field $A_\m$, three Majorana gaugini $\l^I$, $I=1,2,3$, three real scalars $\s^I$, three real auxiliary scalars $D^I$, as well as a Majorana fermion $\c$. All fields are in the adjoint representation of the gauge group.\footnote{In this section we use Hermitian generators for the gauge group, while we find it convenient to formulate the anomaly descent in terms of anti Hermitian generators. The two choices are related as $t^a_{\rm h.}=i\,t^a_{\rm a.h.}$.} As we now show, the $\cn=3$ supersymmetric Chern-Simons action provides a supersymmetric inflow mechanism for the flavor anomaly of the $\cn=(3,3)$ vector multiplet in two dimensions, and hence for any $\cn=(p,q)$ vector multiplet with $p,q\leq 3$.

We should clarify at this point that the flavor anomalies for non-chiral theories, such as $\cn=(3,3)$ theories in 2d or $\cn=2$ ones in 4d that we consider below, are somewhat formal and discussed only as non-trivial solutions of the WZ consistency conditions. Their coefficients vanish in all Lagrangian theories, as well as in non-Lagrangian theories obtained through renormalization group flows from Lagrangian ones, due to 't Hooft anomaly matching. More relevant to the present discussion, however, is the fact that such anomalies for non-chiral multiplets appear as an intermediate step for the computation of supersymmetrized flavor anomalies in chiral theories via anomaly inflow.

The supersymmetry transformations of the $\cn=3$ vector multiplet are parameterized by three real Majorana spinors, $\e^I$, and are given by
\bal
\label{susy-N=3-vector}
\d_Q(\e) \s^I=&\;\ve^{I}{}_{JK}\bar\e^J\l^K-\bar\e^I\c,\NO\\
\d_Q(\e) A_\m=&\;\bar\e^I\g_\m\l^I,\NO\\
\d_Q(\e)\l^I_\a=&\;\frac{1}{2}(\g^{\m\n}\e^I)_\a F_{\m\n}+\ve^{I}{}_{JK}\e^J_\a D^K+\ve^{I}{}_{JK}(\g^\m\e^J)_\a\cd_\m \s^K-i[\s^I,\s_J]\e^J_\a,\NO\\
\d_Q(\e) \c_\a=&\;-\e^I_\a D_I+(\g^\m\e^I)_\a\cd_\m \s_I-\frac{i}{2}\ve_{IJK}[\s^J,\s^K]\e_\a^I,\\
\d_Q(\e) D^I=&\;\ve^{I}{}_{JK}\bar\e^J\g^\m\cd_\m\l^K+\bar\e^I\g^\m\cd_\m\c-i[\bar\e^J\l_J,\s^I]+i[\bar\e^I\l^J+\bar\e^J\l^I,\s_J]-i\ve^{I}{}_{JK}[\bar\e^J\c,\s^K],\NO
\eal
where we follow the spinor conventions of \cite{Freedman:2012zz} and we have suppressed the gauge group indices. The gauge-covariant derivative and field strength are respectively
\be
\cd_\m=\pa_\m-i[A_\m,\;\;],
\ee
\be
F_{\m\n}=\pa_\m A_\n-\pa_\n A_\m-i[A_\m,A_\n].
\ee
Moreover, the indices $I,J,K$ are raised and lowered with the Kronecker delta $\d^{IJ}$, $\d_{IJ}$. Notice that $\ve^{IJK}=\ve_{IJK}$ denotes the Levi-Civita symbol in $\bb R^3$ spanned by the $I,J,K$ indices, while $\ve^{\m\n\r}$ is the Levi-Civita symbol in $\bb R^{1,2}$.

The $\cn=2$ and $\cn=1$ vector multiplets can be obtained from the $\cn=3$ one by setting specific components to zero. The resulting non-zero components for these multiplets are
\bebx\nonumber
\begin{tabular}{c|c|c}
& SUSY parameters & Non-zero components\\\hline
\rule{0cm}{.7cm}$\cn=2$ & $\e^1$, $\e^2$ & $A_\m$, $\s_3$, $D_3$, $\l_1$, $\l_2$\\
\rule{0cm}{.5cm}$\cn=1$ & $\e^1$ & $A_\m$, $\l_1$
\end{tabular}
\eebx
It is straightforward to check that these are consistent truncations of the off-shell supersymmetry transformations \eqref{susy-N=3-vector}.

Using the 3d identities
\be
\d^I_{[K}\,\ve_{L]}{}^{PQ}+\d^{I[P}\,\ve^{Q]}{}_{KL}=0,\qquad \d^I_{[K}\,\ve_{L]}{}^{PQ}+\d^{P}_{[K}\,\ve_{L]}{}^{QI}+\d^{Q}_{[K}\,\ve_{L]}{}^{IP}=0,
\ee
one can show that, together with rigid translations $\d_D(\x) = \x^\m\pa_\m$ and gauge transformations, which act on the vector multiplet fields as $\d_G(\vth) A_\m=\cd_\m\vth$, and $\d_G(\vth)=i[\vth,\;\;]$ for all other fields, the transformations \eqref{susy-N=3-vector} close off-shell and obey the algebra
\be\label{3d-N=3-algebra}
[\d_Q(\e_1),\d_Q(\e_2)]=\d_D(\x)+\d_G(\vth),
\ee
with the composite translation and gauge parameters given respectively by
\be
\x^\m=2\bar\e_2^I\g^\m\e_{1I},\qquad \vth=-\x^\m A_\m-2\ve_{IJK}(\bar\e_2^I\e_1^J)\s^K.
\ee
Notice that $\x^\m$ is constant (as required) while $\vth$ is not due to the explicit field dependence.

The $\cn=3$ supersymmetric Chern-Simons Lagrangian is \cite{Kao:1995gf}
\be\label{d=3-N=3-CS}
\cl_{\rm CS}=\frac{k}{4\pi}\,\tr\Big(\k\,\ve^{\m\n\r}\Big(A_\m\pa_\n A_\r-\frac{2i}{3}A_\m A_\n A_\r\Big)-\bar\l^I\l_I+\bar\c\c-2\s^I D_I+\frac{i}{3}\ve_{IJK}\s^I[\s^J,\s^K]\Big),
\ee
where $k$ is the Chern-Simons level, while $\k=\pm1$ parameterizes a choice in the 3d Clifford algebra through the identity
\be\label{3d-gamma-convention}
\g^{\m\n}=-\k\;\ve^{\m\n\r}\g_\r,\qquad \k=\pm1.
\ee

A generic variation of the Chern-Simons Lagrangian \eqref{d=3-N=3-CS} takes the form
\bal
\d\cl_{\rm CS}=&\;\frac{k}{4\pi}\,\tr\Big(\k\,\ve^{\m\n\r}\big(\pa_\r(A_\n\d A_\m)+\d A_\m F_{\n\r}\big)-2\bar\l^I\d\l_I+2\bar\c\d\c-2\s^I\d D_I\NO\\
&-\d \s^I\big(2D_I-i\ve_{IJK}[\s^J,\s^K]\big)\Big).
\eal
Specializing this variation to gauge and $\cn=3$ supersymmetry transformations, we find that in both cases \eqref{d=3-N=3-CS} is invariant up to a total derivative term, namely
\bebx
\label{d=3-N=3-inflow}
\begin{aligned}
\d_{G}(\vth)\cl_{\rm CS}=&\;\frac{k}{4\pi}\k\,\ve^{\m\n\r}\,\pa_\m\tr(\vth\, \pa_\n A_\r),\\
\d_Q(\e)\cl_{\rm CS}
=&\;\frac{k}{4\pi}\,\pa_\m\tr\big(\k\,\ve^{\m\n\r}\d_Q(\e) A_\n A_\r-2\s^I(\ve_{IJK}\bar\e^J\g^\m\l^K+\bar\e_I\g^\m\c)\big).
\end{aligned}
\eebx
The boundary term resulting from a gauge transformation matches the usual non-Abelian anomaly, i.e. the bosonic part of the 2d flavor anomaly. As we now discuss, the boundary term arising from supersymmetry transformations corresponds to its $(3,3)$ supersymmetric completion, i.e. the $(3,3)$ supersymmetry completion of the anomaly in the WZ gauge, which follows from the WZ consistency conditions.

Upon dimensional reduction to two dimensions, the 3d $\cn=3$ vector multiplet reduces to the 2d $\cn=(3,3)$ vector multiplet (see e.g. \cite{Shima:2006eq} for the Abelian case). This has the same field content as the 3d $\cn=3$ vector multiplet, except that the 3d gauge field gives rise to an extra scalar: $A_\m=(A_{\hat\m},\f)$, where $\hat\m=0,1$. The supersymmetry transformations of the 2d $\cn=(3,3)$ vector multiplet follow from the 3d $\cn=3$ transformations in \eqref{susy-N=3-vector}, namely
\bal
\label{susy-N=3-vector-2D}
\d_Q(\e) \s^I=&\;\ve^{I}{}_{JK}\bar\e^J\l^K-\bar\e^I\c,\NO\\
\d_Q(\e) A_{\hat\m}=&\;\bar\e^I\g_{\hat\m}\l^I,\NO\\
\d_Q(\e) \f=&\;\k\,\bar\e^I\g_*\l^I,\NO\\
\d_Q(\e)\l^I_\a=&\;\frac{1}{2}(\g^{\hat\m\hat\n}\e^I)_\a F_{\hat\m\hat\n}+\k(\g^{\hat\m}\g_*\e^I)_\a \cd_{\hat\m}\f+\ve^{I}{}_{JK}\e^J_\a D^K+\ve^{I}{}_{JK}(\g^{\hat\m}\e^J)_\a\cd_{\hat\m} \s^K\NO\\
&\;-i\k\,\ve^{I}{}_{JK}(\g_*\e^J)_\a[\f,\s^K]-i[\s^I,\s_J]\e^J_\a,\NO\\
\d_Q(\e) \c_\a=&\;-\e^I_\a D_I+(\g^{\hat\m}\e^I)_\a\cd_{\hat\m} \s_I-i\k\,(\g_*\e^I)_\a[\f, \s_I]-\frac{i}{2}\ve_{IJK}[\s^J,\s^K]\e_\a^I,\NO\\
\d_Q(\e) D^I=&\;\ve^{I}{}_{JK}\bar\e^J\g^{\hat\m}\cd_{\hat\m}\l^K-i\k\,\ve^{I}{}_{JK}\bar\e^J\g_*[\f,\l^K]+\bar\e^I\g^{\hat\m}\cd_{\hat\m}\c-i\k\,\bar\e^I\g_*[\f,\c]\NO\\
&\;-i[\bar\e^J\l_J,\s^I]+i[\bar\e^I\l^J+\bar\e^J\l^I,\s_J]-i\ve^{I}{}_{JK}[\bar\e^J\c,\s^K],
\eal
where $\g_*\equiv \k\, \g^2=\k\, \g_2$ is the chirality matrix in two dimensions (also denoted by $\g_3$; see \cite{Freedman:2012zz}). These satisfy the algebra \eqref{3d-N=3-algebra} with 2d parameters
\be
\x^{\hat\m}=2\bar\e_2^I\g^{\hat\m}\e_{1I},\qquad \vth=-\x^{\hat\m} A_{\hat\m}-2\k\,\bar\e_2^I\g_3\e_{1I}\f-2\ve_{IJK}(\bar\e_2^I\e_1^J)\s^K.
\ee

When expressed in terms of the 2d vector multiplet, the symmetry variations \eqref{d=3-N=3-inflow} of the $\cn=3$ Chern-Simons action provide -- by construction -- a solution of the WZ consistency conditions for the 2d $(3,3)$ symmetry algebra. We therefore conclude that the $(3,3)$ flavor anomaly in two dimensions takes the form
\bebx
\label{2d-N=(3,3)-anomaly}
\begin{aligned}
\d_{G}(\vth) W=&\;\frac{k}{4\pi}\ve^{\hat\n\hat\r}\int d^2x\,\tr(\vth\, \pa_{\hat\n} A_{\hat\r}),\\
\d_Q(\e) W=&\;\frac{k}{4\pi}\int d^2x\,\tr\big(\ve^{\hat\n\hat\r}\d_Q(\e) A_{\hat\n} A_{\hat\r}-2\k\,\s^I(\ve_{IJK}\bar\e^J\g_*\l^K+\bar\e_I\g_*\c)\big),
\end{aligned}
\eebx
where $\ve_{(2)}^{\hat\m\hat\n}\equiv\k\,\ve_{(3)}^{2\hat\m\hat\n}$. This generalizes since long known results for 2d gauge anomalies for theories with less supersymmetry \cite{Itoyama:1985ni,Hwang:1985tm,Kaiser:1988zg}. The $(p,q)$ flavor anomaly for any $p,q\leq 3$ can be obtained from the $(3,3)$ anomaly in \eqref{2d-N=(3,3)-anomaly} by a suitable truncation of the vector multiplet.

\subsubsection{5d to 4d Anomaly Inflow}

$\cn\leq 2$ flavor symmetry in four dimensions presents another example of supersymmetric anomaly inflow. The $\cn=1$ off-shell gauge multiplet in five dimensions  \cite{Kugo:2000af,Cortes:2003zd,Freedman:2012zz,Kuzenko:2013rna,Gall:2018ogw}\footnote{To avoid cluttering the notation we use $\m,\n,\r,\ldots$ to denote both 4d and 5d spacetime indices, since the distinction should be clear from the context.} comprises a gauge field $A_\m$, a symplectic Majorana spinor $\l^i$ that transforms as a doublet of the $SU(2)$ $R$-symmetry group, a real scalar $\s$, and an auxiliary real symmetric tensor $Y^{ij}=Y^{ji}$, all in the adjoint representation of the gauge group. Following again the conventions of \cite{Freedman:2012zz}, the supersymmetry transformations are
\bal\label{5d-N=2-vector-susy-trans}
&\d_Q(\e) A_\m^a=\tfrac12\bar\e^i\g_\m\l^a_i,\quad \d_Q(\e)\s^a=\tfrac{i}{2}\bar\e^i\l^a_i,\NO\\
&\d_Q(\e) Y^{ija}=-\tfrac12\bar\e^{(i}\big(\slash{\hspace{-1pt}\cd}\l^{j)a}+if_{bc}{}^a\s^b\l^{j)c}\big),\NO\\
&\d_Q(\e)\l^{ia}=-\tfrac14\g^{\m\n}F_{\m\n}^a\e^i-\tfrac{i}{2}\slash{\hspace{-1pt}\cd}\s^a\e^i-Y^{ija}\e_j,
\eal
where the covariant derivative and field strength of the gauge field are given by
\be
\cd_\m\f^{a}\equiv(\cd_\m\f)^a=\pa_\m\f^{a}+f_{bc}{}^aA^b_\m\f^{c},\qquad \f^a=\text{any field},
\ee
\be
F^a_{\m\n}=\pa_\m A^a_\n-\pa_\n A_\m^a+f_{bc}{}^aA_\m^b A_\n^c.
\ee
Together with rigid translations, $\d_D(\x)=\x^\m\pa_\m$ and gauge transformations that act as $\d_G(\vth)A_\m^a=(\cd_\m\vth)^a=\pa_\m\vth^a+f_{bc}{}^aA_\m^b\vth^c$ and $\d_G(\vth)\f^a=f_{bc}{}^a\f^b\vth^c$ on all other components of the multiplet,  the supersymmetry transformations \eqref{5d-N=2-vector-susy-trans} satisfy the algebra
\be
[\d_Q(\e_1),\d_Q(\e_2)]=\d_D(\x)+\d_G(\vth),\qquad \x^\m=\tfrac12\bar\e_2^i\g^\m\e_{1i},\quad\vth^a=-\x^\m A_\m^a-\tfrac{i}{2}\bar\e^i_2\e_{1i}\s^a.
\ee

An important subtlety in five dimensions is that there exists no pure Chern-Simons action for the vector multiplet. $\cn=2$ supersymmetric Lagrangians are specified by a prepotential $\cf(\s)$, which couples the Chern-Simons and Yang-Mills parts of the action. Although, a pure Yang-Mills Lagrangian is obtained from a quadratic prepotential, a supersymmetric Chern-Simons Lagrangian requires a cubic prepotential and necessarily contains a Yang-Mills part. In particular, the supersymmetric Chern-Simons Lagrangian takes the form
\bal
\label{5dCS}
\cl_{\rm CS}=&\;\big(-\tfrac14F^a_{\m\n}F^{b\m\n}-\tfrac12\bar\l^{ia}\slash{\hspace{-1pt}\cd}\l_i^b-\tfrac12\cd_\m\s^a\cd^\m\s^b+Y^a_{ij}Y^{ijb}\big)\cf_{ab}\NO\\
&\;+\Big(\tfrac{\k}{24}\ve^{\m\n\r\s\t}A_\m^a\Big( F_{\n\r}^bF_{\s\t}^c+f_{de}{}^bA_\n^dA_\r^e\big(-\tfrac12F^c_{\s\t}+\tfrac{1}{10}f_{fg}{}^cA^f_\s A^g_\t\big)\Big)\NO\\
&\;-\tfrac{i}{8}\bar\l^{ia}\g^{\m\n}F_{\m\n}^b\l_i^c-\tfrac{i}{2}\bar\l^{ia}\l^{jb}Y^c_{ij}+\tfrac{i}{4}f_{de}{}^c\s^a\s^b\bar\l^{id}\l^e_i\Big)\cf_{abc},
\eal
where again $\k=\pm 1$ parameterizes a choice in the representation of the Clifford algebra in five dimensions through the relation $\g^{\m\n\r\s\t}=-i\k\,\ve^{\m\n\r\s\t}$. Moreover, $\cf_{ab}$, $\cf_{abc}$ denote respectively the second and third derivatives of the prepotential, $\cf$, which we take to be
\be
\label{prepotential}
\cf(\s)=\frac{k}{48\p^2}d_{abc}\s^a\s^b\s^c,
\ee
where $d_{abc}=\tr(t_a\{t_b,t_c\})$ is the completely symmetric rank-3 invariant tensor on the Lie algebra of the gauge group and $k$ is the gauge/flavor anomaly coefficient that depends on the microscopic theory.

The prepotential \eqref{prepotential} is chosen such that the gauge transformation of the Chern-Simons action \eqref{5dCS} coincides with the bosonic part of the consistent gauge/flavor anomaly in four dimensions upon the identification $\ve_{(4)}^{\m\n\r\s}\equiv \k\,\ve_{(5)}^{4\m\n\r\s}$, namely
\bebx
\label{5dCS-gauge-variation}
\d_{G}(\vth) \cl_{\rm CS}=\frac{k}{48\p^2}\int d^4x\,d_{abc}\vth^a\ve^{\m\n\r\s}\pa_\m\big(A_\n^b\pa_\r A_\s^c+\tfrac14f_{de}{}^cA_\n^bA_\r^dA_\s^e\big).
\eebx
The supersymmetry transformation of the Chern-Simons Lagrangian \eqref{5dCS} is also a total derivative, which we determine next. By construction, the two symmetry transformations of the 5d supersymmetric Chern-Simons action provide a solution of the WZ consistency conditions for the $\cn=2$ symmetry algebra in four dimensions, and therefore determine the $\cn=2$ supersymmetric flavor/anomaly.

A lengthy computation using several identities for 5d symplectic Majorana spinors (see Appendix A of \cite{Cortes:2003zd}) and the Lie algebra relation
\be
d_{c(ab}f_{e)d}{}^c=0,
\ee
determines that the supersymmetry variation of \eqref{5dCS} takes the form
\bebx
\label{5dCS-susy-variation}
\begin{aligned}
\d_Q(\e)\cl_{\rm CS}=&\;\pa_\m\Big[\big(-\tfrac14(\bar\e^i\g_\n\l_i^a)F^{b\m\n}-\tfrac{i}{4}(\bar\e^i\l_i^a)\cd^\m\s^b-\tfrac12(\bar\e_i\g^\m\l^a_j)Y^{ijb}\\
&-\tfrac18(\bar\e^i\g^{\m\r\s}\l_i^a)F^b_{\r\s}-\tfrac{i}{4}(\bar\e^i\g^{\m\n}\l_i^a)\cd_\n\s^b\big)\cf_{ab}\\
&+\tfrac{\k}{6}\ve^{\m\n\r\s\t}\Big(\d_Q(\e) A^a_\n\big(2A_\r^b\pa_\s A_\t^c+\tfrac34 f_{de}{}^bA_\r^cA_\s^dA_\t^e+\tfrac{1}{8}\bar\l^{ib}\g_{\r\s\t}\l_i^c\big)\\
&+\tfrac{1}{32}(\bar\e^i\g_{\n\r}\l_i^a)(\bar\l^{jb}\g_{\s\t}\l_j^c)\Big)\cf_{abc}\Big].
\end{aligned}
\eebx
Once expressed in terms of the 4d $\cn=2$ multiplet fields arising from the dimensional reduction of the 5d $\cn=1$ vector multiplet, this variation coincides with the supersymmetric completion of the $\cn=2$ gauge/flavor anomaly in four dimensions.

The 4d $\cn=2$ vector multiplet possesses the same field content as the corresponding 5d multiplet, except that the components $A_4^a$ of the 5d gauge field combine with the scalars $\s^a$ into a complex scalar: $X^a=\tfrac12(A_4^a-i\s^a)$. Moreover, the 5d symplectic Majorana gaugino reduces to an $SU(2)$ doublet of either chiral or Majorana gauginos in four dimensions. Following \cite{Freedman:2012zz}, we decompose the 5d gaugino and supersymmetry parameter in terms of chiral spinors in four dimensions as (see appendix 20.B in \cite{Freedman:2012zz})
\bal\label{5d4d-spinor-decomposition}
&\l_i^a=-\l_{(4)}^{ja}\ve_{ji}+\l_{(4)i}^a,\quad \bar\l_i^a=\bar\l_{(4)}^{ja}\ve_{ji}+\bar\l_{(4)i}^a,\qquad
\l_{(4)i}^a=P_L\l_{(4)i}^a,\quad \l_{(4)}^{ia}=P_R\l_{(4)}^{ia},\NO\\
\rule{0cm}{.7cm}&\e_i=\e_{(4)}^{j}\ve_{ji}+\e_{(4)i},\quad \bar\e_i=\bar\e_{(4)}^{j}\ve_{ji}-\bar\e_{(4)i},\qquad\e_{(4)i}=P_R\e_{(4)i},\quad \e_{(4)}^{i}=P_L\e_{(4)}^{i},
\eal
where $P_L=\tfrac12(1-\k\g_*)$, $P_R=\tfrac12(1+\k\g_*)$ are the 4d chirality projectors and $\g_*=-\k\g^4$. In particular, $\e_{(4)}^{i}$, $\e_{(4)i}$ and $\l_{(4)}^{ia}$, $\l_{(4)i}^a$ are charge conjugate pairs so that $\e_{(4)}^{i}+\e_{(4)i}$ and $\l_{(4)}^{ia}+\l_{(4)i}^a$ are Majorana.

Inserting this decomposition in \eqref{5d-N=2-vector-susy-trans} and dropping the subscript $(4)$ leads to the 4d $\cn=2$ supersymmetry transformations
\bal\label{4d-N=2-susy-transformations}
\d_Q(\e) X^a=&\;\tfrac12\bar\e^i\l_{i}^a,\NO\\
\d_Q(\e) \l_{i}^a=&\;\slash{\hskip-1pt\cd} X^a\e_i+\tfrac14\g^{\m\n}F_{\m\n}^a\ve_{ij}\e^j+Y_{ij}^a\e^j+X^b\bar X^cf_{bc}{}^a\ve_{ij}\e^j,\NO\\
\d_Q(\e) A_\m^a=&\;\tfrac12 \ve^{ij}\bar\e_i\g_\m\l_j^a+\tfrac12 \ve_{ij}\bar\e^i\g_\m\l^{ja},\NO\\
\d_Q(\e) Y^{ija}=&\;\tfrac12\bar\e^{(i}\slash{\hspace{-1pt}\cd}\l^{j)a}+f_{bc}{}^a X^b\ve^{(ik}\bar\e_k\l^{j)c}+\tfrac12\ve^{(ik}\ve^{j)l}\bar\e_k\slash{\hspace{-1pt}\cd}\l_l^a-f_{bc}{}^a\bar X^b\bar\e^{(i}\l_l^c\ve^{j)l},
\eal
where now $\m,\n=0,1,2,3$. Together with gauge transformations and translations, these satisfy the same algebra as in five dimensions, but with composite parameters
\be
\x^\m=\tfrac12\bar\e_2^i\g^\m\e_{1i}+\text{h.c.},\quad\vth^a=-\x^\m A_\m^a+X^a\ve^{ij}\bar\e_{2i}\e_{1j}+\bar X^a\ve_{ij}\bar\e^i_2\e^j_1.
\ee

We can now evaluate the symmetry transformations \eqref{5dCS-gauge-variation} and \eqref{5dCS-susy-variation} of the 5d Chern-Simons Lagrangian in terms of the 4d $\cn=2$ vector multiplet components in order to obtain the $\cn=2$ supersymmetric gauge/flavor anomaly in four dimensions:
\bebx
\label{5d-N=2-flavor-anomaly}
\begin{aligned}
\d_{G}(\vth) W=&\;\frac{k}{48\p^2}\int d^4x\,d_{abc}\vth^a\ve^{\m\n\r\s}\pa_\m\big(A_\n^b\pa_\r A_\s^c+\tfrac14f_{de}{}^cA_\n^bA_\r^dA_\s^e\big),\\
\rule{.0cm}{.7cm}\d_Q(\e) W=&\;\frac{k}{16\p^2}\int d^4x\,d_{abc}\Big[i(X^c-\bar X^c)\Big(\ve_{ij}(\bar\e^i\g^\n\l^{ja})\cd_\n X^b+\ve^{ij}(\bar\e_i\g^\n\l_j^{a})\cd_\n \bar X^b\\
&\hspace{-1.5cm}-(\bar\e^i\l_i^a-\bar\e_i\l^{ia})f_{de}{}^bX^d\bar X^e+(\bar\e^j\l^a_i+\bar\e_i\l^{ja})Y^{i}{}^b_j-\tfrac{1}{4}(\bar\e^i\g^{\r\s}\l_i^a+\bar\e_i\g^{\r\s}\l^{ia})F^b_{\r\s}\Big)\\
&\hspace{-1.5cm}+\tfrac{1}{3}\ve^{\n\r\s\t}\Big(\d_Q(\e) A^a_\n\big(2A_\r^b\pa_\s A_\t^c+\tfrac34 f_{de}{}^bA_\r^cA_\s^dA_\t^e+\tfrac{1}{8}\bar\l^{ib}\g_{\r\s\t}\l_i^c+\tfrac{1}{8}\bar\l_i^{b}\g_{\r\s\t}\l^{ic}\big)\\
&\hspace{-1.5cm}+\tfrac{1}{32}(\bar\e^i\g_{\n\r}\l_i^a-\bar\e_i\g_{\n\r}\l^{ia})(\ve_{kl}\bar\l^{kb}\g_{\s\t}\l^{lc}+\ve^{kl}\bar\l_k^{b}\g_{\s\t}\l_l^{c})\Big)\Big].
\end{aligned}
\eebx
This generalizes well known results for the $\cn=1$ gauge/flavor anomaly, which we can easily recover from \eqref{5d-N=2-flavor-anomaly} through a truncation of the $\cn=2$ multiplet.

The 4d $\cn=1$ vector multiplet with a Majorana gaugino and the corresponding supersymmetry transformations can be obtained from the $\cn=2$ multiplet by setting
\bal\label{4d-N=2-N=1-truncation}
&\e_1=\e^1=0,\quad \l^a_2=\l^{2a}=0,\quad X^a=0,\quad Y^{11a}=Y^{22a}=0,\NO\\
&\e\equiv\e_2+\e^2,\quad \l^a\equiv\l_1^a+\l^{1a},\quad D^a\equiv -2i\k Y^{12a}.
\eal
Inserting these in \eqref{4d-N=2-susy-transformations} results in the $\cn=1$ supersymmetry transformations
\be\label{4d-N=1-susy}
\d_Q(\e)A_\m^a=-\tfrac12\bar\e\g_\m\l^a,\quad
\d_Q(\e)\l^a=\big(\tfrac14\g^{\m\n}F_{\m\n}^a+\tfrac{i}{2}\g_* D^a\big)\e,\quad
\d_Q(\e)D^a=\tfrac{i}{2}\bar\e\g_*\g^\m\cd_\m\l^a,
\ee
which satisfy the algebra (the subscripts 1, 2 here should not to be confused with the $SU(2)$ indices of the $\cn=2$ multiplet)
\be
[\d_Q(\e_1),\d_Q(\e_2)]=\d_D\big(\tfrac12\bar\e_2\g^\m\e_1\big)+\d_G\big(-\tfrac12\bar\e_2\g^\m\e_1A_\m\big).
\ee

Evaluating the transformations \eqref{5d-N=2-flavor-anomaly} on the truncated multiplet \eqref{4d-N=2-N=1-truncation} results in the $\cn=1$ supersymmetric gauge/flavor anomaly in four dimensions \cite{Itoyama:1985qi,Guadagnini:1985ea,Itoyama:1985ni}
\bebx\label{4dN=1anomalies}
\begin{aligned}
\hspace{-.25cm}\d_{G}(\vth) W=&\;\frac{k}{48\p^2}\int d^4x\,d_{abc}\vth^a\ve^{\m\n\r\s}\pa_\m\big(A_\n^b\pa_\r A_\s^c+\tfrac14f_{de}{}^cA_\n^bA_\r^dA_\s^e\big),\\
\hspace{-.25cm}\d_Q(\e) W=&\;\frac{k}{48\p^2}\int d^4x\,d_{abc}\ve^{\m\n\r\s}\d_Q(\e)A_\m^a\big(2A_\n^b\pa_\r A_\s^c+\tfrac34 f_{de}{}^bA_\n^cA_\r^dA_\s^e+\tfrac{1}{8}\bar\l^b\g_{\n\r\s}\l^c\big).\hspace{-.25cm}
\end{aligned}
\eebx

\subsection{Current Multiplet Anomaly Inflow}

We now turn to anomaly inflow for local supersymmetry, which forms an algebra with diffeomorphisms, local Lorentz transformations and $R$-symmetry. The gauge fields corresponding to these local symmetries comprise an off-shell supergravity multiplet, which couples minimally to the current multiplet containing the stress tensor, the supercurrent and R-current.

For a fixed amount of supersymmetry in a given spacetime dimension there exist different off-shell supergravity multiplets that differ in auxiliary field content and the symmetries they gauge. The gravity multiplet with the minimal auxiliary field content is off-shell conformal supergravity that gauges the entire superconformal group and includes Weyl and S-supersymmetry transformations. Any other off-shell supergravity multiplet with the same amount of supersymmetry can be obtained by coupling specific matter multiplets to conformal supergravity and imposing suitable gauge-fixing conditions. This results in a larger field content relative to conformal supergravity, but less symmetries due to the partial gauge fixing conditions. For example, old minimal \cite{Ferrara:1978em,Stelle:1978ye,Fradkin:1978jq}, new minimal \cite{Akulov:1976ck,Sohnius:1981tp,Sohnius:1982fw,Ferrara:1988qxa} and 16+16 \cite{Girardi:1984vq,Lang:1985xk,Siegel:1986sv} supergravities in four dimensions can all be obtained in this way from $\cn=1$ conformal supergravity \cite{Kaku:1977pa,Kaku:1977rk,Kaku:1978nz,Townsend:1979ki}. These couple respectively to the Ferrara-Zumino, R, and S current multiplets \cite{Komargodski:2010rb}, while conformal supergravity couples to the conformal current multiplet.

Gravitational and Lorentz anomalies exist in $d=4k+2$ dimensions and can be obtained from a Chern-Simons action in $d=4k+3$ dimensions \cite{AlvarezGaume:1983ig,Bardeen:1984pm}. Like flavor anomalies, gravitational and Lorentz anomalies result in an associated Q-supersymmetrized anomaly, which contributes a local term to the divergence of the supercurrent \cite{Tanii:1985wy,Howe:1985uy,Itoyama:1985ni}. For the minimal $\cn=(1,0)$ supergravity in two dimensions, the supersymmetric completion of the gravitational anomaly has also been shown to follow from a supersymmetric gravitational Chern-Simons action in three dimensions \cite{Tanii:1985wy}. Another local contribution to the divergence of the supercurrent arises in all even dimensions in the presence of an $R$-symmetry anomaly \cite{Papadimitriou:2017kzw,An:2019zok,Papadimitriou:2019gel,Katsianis:2019hhg,Papadimitriou:2019yug,Katsianis:2020hzd}. The $R$-symmetry anomaly can be obtained from a Chern-Simons action in $d+1$ dimensions too, and one expects that its supersymmetric completion should follow similarly from a supersymmetric Chern-Simons action.

In this subsection we discuss the gravitational/Lorentz anomaly of the $(p,q)$ conformal current multiplet in two dimensions and, in particular, all mixed anomalies it generates. As we will see, there are two effects that control the structure of these anomalies. Firstly, local supersymmetry requires that the gravitational/Lorentz and $R$-symmetry anomalies be considered in tandem, since the underlying algebra mixes the corresponding symmetries. Secondly, both the gravitational/Lorentz and $R$-symmetry anomalies produce mixed anomalies for all other symmetries of the multiplet. Our goal here is to obtain all these anomalies through an inflow mechanism from an off-shell supergravity Chern-Simons action in three dimensions \cite{vanNieuwenhuizen:1985cx,Rocek:1985bk,Cederwall:2011pu,Nishimura:2012jh,Kuzenko:2012ew,Butter:2013goa,Butter:2013rba,Nishimura:2013poa,Kuzenko:2013vha}.

As a side comment, we note that all anomalies of the $(p,q)$ conformal current multiplet in two dimensions may alternatively be obtained holographically from the Chern-Simons action of the (on-shell) $(p,q)$ AdS$_3$ supergravity of Achucarro and Townsend \cite{Achucarro:1987vz,Achucarro:1989gm}, which gauges the supergroup $OSp(p|2;\bb R)\times OSp(q|2;\bb R)$.\footnote{See \cite{Nishimura:1999gg} for an early special case of such a derivation.} Such a holographic calculation would reproduce not only the gravitational/Lorentz anomaly and the resulting mixed anomalies \cite{Banados:2004zt,Solodukhin:2005ns}, but also the Weyl anomaly \cite{Henningson:1998gx} and its supersymmetric completion \cite{Papadimitriou:2017kzw,An:2017ihs}. However, here we are interested specifically in obtaining the current multiplet anomalies through anomaly inflow, and so we focus on the gravitational/Lorentz anomaly.

\subsubsection{3d to 2d Anomaly Inflow}

The maximal off-shell conformal supergravity (Weyl) multiplet is the $\cn=8$ multiplet \cite{Cederwall:2011pu,Nishimura:2012jh}, but a Chern-Simons action is known only for up to $\cn=6$ \cite{Nishimura:2013poa,Kuzenko:2013vha}. The $R$-symmetry group of these multiplets is $SO(\cn)$ with the exception of the $\cn=6$ multiplet, in which case it is enhanced to $SO(6)\times U(1)\cong U(4)$. Upon dimensional reduction to two dimensions, the 3d gravity multiplets reduce to $(\cn,\cn)$ off-shell conformal supergravity in two dimensions, which gauges the $OSp(\cn|2;\bb R)\times OSp(\cn|2;\bb R)$ conformal group. A peculiarity of 2d supergravity is that a single $SO(\cn)$ gauge field gauges both left and right copies of the $SO(\cn)_L\times SO(\cn)_R$ $R$-symmetry \cite{Gates:1988ey}.

We follow the component formulation of \cite{Nishimura:2013poa} and focus on the $\cn=6$ Weyl multiplet since it is the maximal one for which a Chern-Simons action is known. All multiplets with less supersymmetry can be obtained by suitable truncations, as shown in table \ref{2dCSGtruncations}. The field content of the off-shell $\cn=6$ Weyl multiplet consists of the dreibein $e_\m{}^a$, six Majorana gravitini $\j^I_\m$, $I=1,\cdots,6$, the $SO(6)\times U(1)$ gauge fields, respectively $B_\m^{IJ}$ and $B_\m$, as well as two sets of auxiliary Majorana spinors, $\l^{IJK}$ and $\l^I$, and two sets of auxiliary scalars $E^{IJ}$ and $D^{IJ}$. The $SO(6)$ indices $I,J,K=1,\cdots,6$ in all fields are totally antisymmetrized. The Weyl weight and $R$-symmetry representation of all components are given in table \ref{2dCSGcontent}.
\begin{table}
\bebx\nonumber
\begin{tabular}{c|c|c|c|c|c|c|c|c}
Field & $e_\m{}^a$ & $\j_\m^I$ & $B_\m^{IJ}$ & $B_\m$ & $\l^{IJK}$ & $\l^I$ & $E^{IJ}$ & $D^{IJ}$ \raisebox{-.2cm}{\rule{0cm}{.65cm}}\\\hline
Weyl weight & $1$ & $\tfrac12$ & $0$ & $0$ & $-\tfrac32$ & $-\tfrac32$ & $-1$ & $-2$\raisebox{-.2cm}{\rule{0cm}{.65cm}}\\\hline
$SU(4)$ representation & $\bf 1$ & $\bf 6$ & $\bf 15$ & $\bf 1$ & $\bf 20$ & $\bf 6$ & $\bf 15$ & $\bf 15$ \raisebox{-.2cm}{\rule{0cm}{.65cm}}\\\hline
$U(1)$ charge & $0$ & $0$ & $0$ & $0$ & $0$ & $0$ & $0$ & $0$\raisebox{-.2cm}{\rule{0cm}{.65cm}}
\end{tabular}
\eebx
\captionof{table}{The field content of the off-shell $\cn=6$ conformal supergravity field content. All $SO(6)$ indices $I,J,K=1,\cdots,6$ are totally antisymmetrized. Notice that all components are neutral under the $U(1)$ factor of the $R$-symmetry group.}
\label{2dCSGcontent}
\end{table}

The local symmetries of the $\cn=6$ Weyl multiplet comprise diffeomorphisms, parameterized by the infinitesimal vector field $\x^\m(x)$, local Lorentz transformations, $\l^{ab}(x)=-\l^{ba}(x)$, Weyl rescalings, $\s(x)$, $SO(6)\times U(1)$ $R$-symmetry transformations parameterized by $\th^{IJ}(x)=-\th^{JI}(x)$ and $\th(x)$, as well as Q- and S-supersymmetry transformations, parameterized respectively by the local Grassmann-valued Majorana parameters $\e^I(x)$ and $\h^I(x)$. The infinitesimal transformations of all fields under the bosonic symmetries take the form
\bal\label{CSG-bosonc-transformations}
\d_Be_\m{}^a=&\;\x^\n\pa_\n e_\m{}^a+\pa_\m\x^\n e_\n{}^a-\l^a{}_b e_\m{}^b+\s e_\m{}^a,\NO\\
\d_B\j_\m^I=&\;\x^\n\pa_\n \j_\m^I+\pa_\m\x^\n \j_\n^I-\tfrac14\l_{ab}\g^{ab} \j_\m^I+\tfrac12\s \j_\m^I-\th^{IJ}\j_\m^J,\NO\\
\d_BB_\m^{IJ}=&\;\x^\n\pa_\n B_\m^{IJ}+\pa_\m\x^\n B_\n^{IJ}+\pa_\m\th^{IJ}+B^{IK}\th_\m^{KJ}+B^{JK}\th_\m^{IK},\NO\\
\d_BB_\m=&\;\x^\n\pa_\n B_\m+\pa_\m\x^\n B_\n+\pa_\m\th,\NO\\
\d_B\l^{IJK}=&\;\x^\n\pa_\n \l^{IJK}-\tfrac14\l_{ab}\g^{ab}\l^{IJK}-\tfrac32\s \l^{IJK}-\th^{IL}\l^{LJK}-\th^{JL}\l^{ILK}-\th^{KL}\l^{IJL},\NO\\
\d_B\l^I=&\;\x^\n\pa_\n \l^{I}-\tfrac14\l_{ab}\g^{ab}\l^{I}-\tfrac32\s \l^{I}-\th^{IL}\l^{LJK},\NO\\
\d_BE^{IJ}=&\;\x^\n\pa_\n E^{IJ}-\s E^{IJ}-\th^{IK}E^{KJ}-\th^{JK}E^{IK},\NO\\
\d_BD^{IJ}=&\;\x^\n\pa_\n D^{IJ}-2 \s D^{IJ}-\th^{IK}D^{KJ}-\th^{JK}D^{IK},
\eal
where we have used the shorthand notation $\d_B\equiv \d_D(\x)+\d_L(\l)+\d_W(\s)+\d_R(\th)$.

Under Q-supersymmetry the components of the Weyl multiplet transform as
\bal\label{CSG-Q-transformations}
\d_Q(\e) e_\m{}^a=&\;\tfrac14\bar\e^I\g^a\j_\m^I,\qquad \d_Q(\e)\j_\m^I=\cd_\m\e^I,\NO\\
\d_Q(\e) B_{\m}^{IJ}=&\;-\bar\e^{[I}\j_{\m+}^{J]}+\tfrac{1}{2\sqrt{2}}\bar\e^K\g_\m\l^{IJK}+\tfrac{1}{4\sqrt{2}}\ve^{IJKLMN}\bar\e^K\j_\m^L E^{MN},\NO\\
\d_Q(\e) B_{\m}=&\;\tfrac{1}{2\sqrt{2}}\bar\e^I\g_\m\l^I-\tfrac{1}{2\sqrt{2}}\bar\e^I\j_\m^J E^{IJ},\NO\\
\d_Q(\e) \l^{IJK}=&\;-\tfrac{3}{4\sqrt{2}}\g^{\m\n}\e^{[I}\Hat G_{\m\n}^{JK]}+\tfrac12\ve^{IJKLMN}\e^L D^{MN}\NO\\
&+\tfrac14\ve^{IJKLMN}\g^\m\e^L\Hat\cd_\m E^{MN}-\tfrac{3}{\sqrt{2}}\e^L E^{[IJ}E^{KL]},\NO\\
\d_Q(\e) \l^{I}=&\;-\tfrac{1}{4\sqrt{2}}\g^{\m\n}\e^I\Hat G_{\m\n}+\e^J D^{IJ}-\tfrac12\g^\m\e^J\Hat\cd_\m E^{IJ}+\tfrac{1}{8\sqrt{2}}\ve^{IJKLMN}\e^J E^{KL}E^{MN},\NO\\
\d_Q(\e) E^{IJ}=&\;\tfrac12\bar\e^{[I}\l^{J]}-\tfrac{1}{24}\ve^{IJKLMN}\bar\e^K\l^{LMN},\NO\\ \d_Q(\e) D^{IJ}=&\;\tfrac12\bar\e^{[I}\l_+^{J]}+\tfrac{1}{24}\ve^{IJKLMN}\bar\e^K\l_+^{LMN},
\eal
while S-supersymmetry acts as
\bal\label{CSG-S-transformations}
\d_S(\h) e_\m{}^a=&\;0,\qquad \d_S(\h)\j_\m^I=\g_\m\h^I,\qquad \d_S(\h)B_\m^{IJ}=\tfrac12\bar\h^{[I}\j_\m^{J]},\NO\\
\d_S(\h)B_\m=&\;0,\qquad \d_S(\h)\l^{IJK}=-\frac12\ve^{IJKLMN}\h^L E^{MN},\qquad \d_S(\h)\l^I=\h^J E^{IJ},\NO\\
\d_S(\h) E^{IJ}=&\;0,\qquad \d_S(\h) D^{IJ}=-\tfrac14\bar\h^{[I}\l^{J]}-\tfrac{1}{48}\ve^{IJKLMN}\bar\h^K\l^{LMN}.
\eal

Following \cite{Nishimura:2013poa}, we have introduced the abbreviations
\bal
\j_{\m+}^I=&\;\tfrac14\g^{\r\s}\g_\m\j_{\r\s}^I,\qquad \j_{\m\n}^I=\cd_{[\m}\j_{\n]}^I,\NO\\
\l^{IJK}_+=&\;-\tfrac12\g^\m\Hat\cd_\m\l^{IJK}+\tfrac12\ve^{IJKLMN}\g^\m\j_{\m+}^L E^{MN}-\tfrac{3}{4\sqrt{2}}\ve^{MNPQ[IJ}\l^{K]MN}E^{PQ}+\tfrac{3}{\sqrt{2}}\l^{[I}E^{JK]},\NO\\
\l^I_+=&\;-\tfrac12\g^\m\Hat\cd_\m\l^I-\g^\m\j_{\m+}^JE^{IJ}+\frac{1}{2\sqrt{2}}\l^{IJK}E^{JK},\NO\\
\Hat G_{\m\n}^{IJ}=&\;G_{\m\n}^{IJ}+2\bar\j_{[\m}^{[I}\j_{\n]+}^{J]}-\tfrac{1}{\sqrt{2}}\bar\j^K_{[\m}\g_{\n]}\l^{IJK}-\tfrac{1}{4\sqrt{2}}\ve^{IJKLMN}\bar\j_\m^K\j_\n^L E^{MN},\NO\\
\Hat G_{\m\n}=&\; G_{\m\n}-\tfrac{1}{\sqrt{2}}\bar\j_{[\m}^I\g_{\n]}\l^I+\frac{1}{2\sqrt{2}}\bar\j_\m^I\j_\n^JE^{IJ},\NO\\
\Hat\cd_\m E^{IJ}=&\;\cd_\m E^{IJ}-\tfrac12\bar\j_\m^{[I}\l^{J]}+\tfrac{1}{24}\ve^{IJKLMN}\bar\j_\m^K\l^{LMN},\NO\\
\Hat\cd_\m\l^{IJK}=&\;\cd_\m\l^{IJK}+\tfrac{3}{4\sqrt{2}}\g^{\r\s}\j_\m^{[I}\Hat G_{\r\s}^{JK]}-\tfrac12\ve^{IJKLMN}\j_\m^L D^{MN}-\tfrac14\ve^{IJKLMN}\g^\r\j_\m^L\Hat \cd_\r E^{MN}\NO\\
&+\tfrac{3}{\sqrt{2}}\j_\m^L E^{[IJ}E^{KL]},\\
\Hat\cd_\m\l^I=&\;\cd_\m\l^I+\tfrac{1}{4\sqrt{2}}\g^{\r\s}\j_\m^I\Hat G_{\r\s}-\j_\m^JD^{IJ}+\tfrac12\g^\r\j_\m^J\Hat\cd_\r E^{IJ}-\tfrac{1}{8\sqrt{2}}\ve^{IJKLMN}\j_\m^J E^{KL}E^{MN},\NO
\eal
where $G_{\m\n}^{IJ}$ and $G_{\m\n}$ denote the field strengths of the $R$-symmetry gauge fields
\bal
G_{\m\n}^{IJ}=&\;\pa_\m B_\n^{IJ}-\pa_\n B_\m^{IJ}+B_\m^{IK}B_\n^{KJ}-B_\n^{IK}B_\m^{KJ},\NO\\
G_{\m\n}=&\;\pa_\m B_\n-\pa_\n B_\m,
\eal
and the covariant derivative $\cd_\m$ includes the $R$-symmetry transformation of the fields, e.g.
\be
\cd_\m\e^I=\big(\pa_\m+\tfrac14\Hat\o_{\m ab}\g^{ab}\big)\e^I+B_{\m}^{IJ}\e^J.
\ee

The quantity $\Hat\o_{\m ab}(e,\j)$ is the torsionful spin connection
\be
\Hat\o_{\m ab}(e,\j)=\o_{\m ab}(e)+\tfrac18(\bar\j_a^I\g_\m\j_b^I+\bar\j_\m^I\g_a\j_b^I-\bar\j_\m^I\g_b\j_a^I),
\ee
with $\o_{\m ab}(e)$ denoting the torsion-free connection. In particular,
\be
\cd_\m e_\n{}^a-\cd_\n e_\m{}^a=\tfrac14\bar\j_\m^I\g^a\j_\n^I.
\ee
Moreover, the Riemann curvature of $\Hat\o_{\m ab}(e,\j)$
\be
\Hat R_{\m\n}{}^a{}_b\equiv 2(\pa_{[\m}\Hat\om_{\n]}{}^a{}_b+\Hat\o_{[\m}{}^a{}_c\,\Hat\o_{\n]}{}^c{}_b),
\ee
satisfies the Bianchi identities
\be
\Hat R_{\m\n\r\s}+\Hat R_{\r\m\n\s}+\Hat R_{\n\r\m\s}=\frac32(\Upsilon_{\m\n\r\s}+\Upsilon_{\r\m\n\s}+\Upsilon_{\n\r\m\s}),\qquad \Hat R_{[\m\n]}=-\frac34\Upsilon^\r{}_{\r\m\n},
\ee
where
\be
\Upsilon^\m{}_{\n\r\s}\equiv\bar\j_{[\n}^I\g^\m\j^I_{\r\s]}.
\ee
The key reason for introducing the torsionful spin connection is that it transforms nicely under both Q- and S-supersymmetry, namely
\bal
\d_Q(\e)\Hat\o_{\m ab}=&\;-\tfrac14\bar\e^I(\g_\m\j^I_{ab}+\g_a\j^I_{\m b}-\g_{b}\j^I_{\m a}),\NO\\
\d_S(\h)\Hat\o_{\m ab}=&\;-\tfrac14\bar\h^I(\g_{ab}\j^I_{\m}+e_{\m a}\j^I_{b}-e_{\m b}\j^I_{a}).
\eal

The local transformations of the $\cn=6$ Weyl multiplet close off-shell. In particular, the commutators between the fermionic transformations satisfy \cite{Nishimura:2012jh}
\bal
\big[\d_Q(\e_1),\d_Q(\e_2)\big]=&\;\d_D(\x)+\d_L(\l)+\d_R(\th)+\d_Q(\e')+\d_S(\h'),\NO\\
\big[\d_Q(\e),\d_S(\h)\big]=&\;\d_W(\s)+\d_L(\l')+\d_R(\th')+\d_S(\h''),\NO\\
\big[\d_S(\h_1),\d_S(\h_2)\big]=&\;0,
\eal
where the composite transformation parameters on the r.h.s. are given by
\bal
\x^\m=&\;\tfrac14\bar\e_2^I\g^\m\e_1^I,\qquad \l_{ab}=-\x^\m\Hat\o_{\m ab},\NO\\
\th^{IJ}=&\;-\x^\m B_\m^{IJ}+\tfrac{1}{4\sqrt{2}}\ve^{IJKLMN}\bar\e_2^K\e_1^L E^{MN},\NO\\
\th=&\;-\x^\m B_\m-\tfrac{1}{2\sqrt{2}}\bar\e_2^K\e_1^L E^{KL},\qquad \e'^I=-\x^\m\j_\m^I,\NO\\
\h'^I=&\;\tfrac12\x^\m\g^{\r\s}\g_\m\j_{\r\s}^I-\tfrac{1}{16}\bar\e_2^{[I}\e_1^{J]}\g^{\r\s}\j_{\r\s}^J-\tfrac{1}{2\sqrt{2}}\bar\e^K_2\e_1^L\l^{IKL}-\tfrac{1}{16}\bar\e_2^{(I}\g^\m\e_1^{J)}(2\g^{\r\s}\g_\m+\g_\m\g^{\r\s})\j_{\r\s},\NO\\
\s=&\;-\tfrac14\bar\e^I\h^I,\qquad \l'_{ab}=\tfrac14\bar\e^I\g_{ab}\h^I,\qquad \th'^{IJ}=-\tfrac12\bar\e^{[I}\h^{J]},\NO\\
\th'=&\;=0,\qquad \h''^I=\tfrac18\g^\m\e^I(\bar\h^J\j_\m^J).
\eal

A supersymmetric Chern-Simons action for the $\cn=6$ Weyl multiplet was found in \cite{Nishimura:2013poa,Kuzenko:2013vha}. In the component formulation of \cite{Nishimura:2013poa} it takes the form
\bal
\label{3dCSG}
\cl_{\rm CSG}=&\;\frac{k}{4\p}\Big[\frac12\ve^{\m\n\r}\Big(\Hat\o_\m{}^a{}_b\pa_\n\Hat\o_\r{}^b{}_a+\frac23\Hat\o_\m{}^a{}_b\Hat\o_\n{}^b{}_c\Hat\o_\r{}^c{}_a\Big)+\frac14e\,\bar\j_{\m\n}\g^{\r\s}\g^{\m\n}\j_{\r\s}\NO\\
&-\ve^{\m\n\r}\Big(B_\m^{IJ}\pa_\n B_\r^{JI}+\frac23B_\m^{IJ}B_\n^{JK}B_\r^{KI}\Big)-2\ve^{\m\n\r}B_\m\pa_\n B_\r\NO\\
&+\frac13e\,\bar\l^{IJK}\l^{IJK}-2e\,\bar\l^I\l^I-8e\,D^{IJ}E^{IJ}+\frac{1}{3\sqrt{2}}e\,\ve^{IJKLMN}E^{IJ}E^{KL}E^{MN}\NO\\
&+\frac16e\,\ve^{IJKLMN}\bar\l^{IJK}\g^\m\j_\m^L E^{MN}+2e\,\bar\l^I\g^\m\j_\m^J E^{IJ}\NO\\
&+e\,\bar\j^I_\m\g^{\m\n}\j_\n^J\Big(E^{IK}E^{JK}-\frac14\d^{IJ}E^{KL}E^{KL}\Big)\Big],
\eal
where $e\equiv\det(e_\m{}^a)$ and we have chosen the convention $\k=1$ in \eqref{3d-gamma-convention}. All off-shell Chern-Simons actions for Weyl multiplets with less supersymmetry found earlier \cite{vanNieuwenhuizen:1985cx,Rocek:1985bk,Kuzenko:2012ew,Butter:2013goa,Butter:2013rba} can be obtained by consistently truncating the $\cn=6$ multiplet as indicated in table \ref{2dCSGtruncations}.
\begin{table}
\bebx\NO
\begin{tabular}{l|l}
 & \hskip1.cm Non-zero components\\\hline
\rule{0cm}{.7cm}$\cn=5$ &  $e_\m{^a}$, $\j_\m^I$,  $B_\m^{IJ}$, $\l^{IJK}$, $\l^6$, $E^{I6}$, $D^{I6}$\\
\rule{0cm}{.5cm}$\cn=4$  & $e_\m{^a}$, $\j_\m^I$,  $B_\m^{IJ}$, $\l^{IJK}$, $E^{56}$, $D^{56}$\\
\rule{0cm}{.5cm}$\cn=3$  & $e_\m{^a}$, $\j_\m^I$,  $B_\m^{IJ}$, $\l^{123}$\\
\rule{0cm}{.5cm}$\cn=2$  & $e_\m{^a}$, $\j_\m^I$,  $B_\m^{12}$\\
\rule{0cm}{.5cm}$\cn=1$  &$e_\m{^a}$, $\j_\m^1$
\end{tabular}
\eebx
\captionof{table}{Consistent truncations of the $\cn=6$ Weyl multiplet.}
\label{2dCSGtruncations}
\end{table}

The Chern-Simons Lagrangian \eqref{3dCSG} is invariant under diffeomorphisms tangent to the boundary, but all other symmetries, namely Weyl, Lorentz, $R$-symmetry, Q- and S-supersymmetry, result in a non-vanishing total derivative term. In particular, we find
\bebx\label{3dCSG-trans}
\begin{aligned}
\d_W(\s)\cl_{\rm CSG}=&\;\frac{k}{4\p}\ve^{\m\n\r}\pa_\m\big(\tfrac12\d_W(\s)\Hat\o_{\n}{}^a{}_b\,\Hat\o_\r{}^b{}_a\big)=\frac{k}{4\p}\ve^{\m\n\r}\pa_\m\big(\Hat\o_{\n\r}{}^\s\pa_\s\s\big),\\
\d_L(\l)\cl_{\rm CSG}=&\;\frac{k}{8\p}\ve^{\m\n\r}\pa_\m\big(\l^a{}_b\pa_\n\Hat\o_{\r}{}^b{}_a\big),\\
\d_R(\th)\cl_{\rm CSG}=&\;\frac{k}{4\p}\ve^{\m\n\r}\pa_\m\big(\th^{IJ}\pa_\n B_\r^{IJ}-2\th\pa_\n B_\r\big),\\
\d_Q(\e)\cl_{\rm CSG}=&\;\frac{k}{4\p}\ve^{\m\n\r}\pa_\m\big(\tfrac12\d_Q(\e)\Hat\o_{\n}{}^a{}_b\,\Hat\o_\r{}^b{}_a+\d_Q(\e)B_\n^{IJ}\,B_\r^{IJ}-2\d_Q(\e)B_\n\,B_\r\big)\\
&+\frac{k}{4\p}\pa_\m\big(-\tfrac16 e\,\ve^{IJKLMN}\bar\e^N\g^\m\l^{KLM}E^{IJ}+2e\,\bar\e^I\g^\m\l^JE^{IJ}\\
&+2e\,\bar\e^I\g^{\m\n}\j_\n^J\big(E^{IK}E^{JK}-\tfrac14\d^{IJ}E^{KL}E^{KL}\big)\big),\\
\d_S(\h)\cl_{\rm CSG}=&\;\frac{k}{4\p}\ve^{\m\n\r}\pa_\m\big(\tfrac12\d_S(\h)\Hat\o_{\n}{}^a{}_b\,\Hat\o_\r{}^b{}_a+\d_S(\h)B_\n^{IJ}B_\r^{IJ}+\bar\h^I\psi_{\n\r}^I\big).
\end{aligned}
\eebx
Upon dimensional reduction to two dimensions these boundary terms produce the Lorentz anomaly of the 2d $\cn=(6,6)$ multiplet and all associated mixed anomalies, demonstrating that they are obtainable through an inflow mechanism. The corresponding anomalies for any $\cn=(p,p)$ multiplet with $p\leq 6$ can be obtained by a suitable truncation of the 3d $\cn=6$ multiplet as indicated in table \ref{2dCSGtruncations}. For chiral multiplets with $\cn=(p,q)$, $q<p$, one must first dimensionally reduce the boundary terms obtained from the $\cn=p$ Chern-Simons action to two dimensions to obtain the anomalies for the $\cn=(p,p)$ multiplet, and then further truncate the 2d multiplet to $\cn=(p,q)$.

We should stress that the Weyl anomaly in \eqref{3dCSG-trans} is not the usual 2d Weyl anomaly, but rather a supersymmetric version of the mixed Lorentz-Weyl anomaly \cite{Chamseddine:1992ry}. Moreover, once truncated to $\cn=1$ supersymmetry,  \eqref{3dCSG-trans} reproduces the result of \cite{Tanii:1985wy} for minimal Poincar\'e supergravity. To see this, one must first truncate the $\cn=6$ conformal supergravity multiplet to the $\cn=1$ one as indicated in table \ref{2dCSGtruncations}, and then construct the corresponding $\cn=1$ Poincar\'e supergravity by coupling a compensating chiral multiplet. This is directly analogous to the construction of old minimal supergravity in four dimensions from $\cn=1$ conformal supergravity (see \cite{Katsianis:2020hzd} for a recent review of this construction). The supersymmetry transformation of the Poincar\'e multiplet is a field dependent linear combination of the Q- and S-supersymmetry transformations of conformal supergravity, namely $\d_Q^{\text{Poincar\'e}}(\e)=\d_Q(\e)+\d_S(\h=S\e)$, where $S$ is an auxiliary real scalar field, a component of the compensating chiral multiplet. As a result, the supersymmetric completion of the Lorentz anomaly for minimal Poincar\'e supergravity is the sum of the Q and S anomalies in \eqref{3dCSG-trans}, with all R-symmetry gauge fields and (conformal supergravity) auxiliary fields set to zero.

\section{Anomaly Descent with Mismatching Ghost}
\label{sec:descent}

We have seen earlier that, for diffeomorphisms, the standard anomaly
descent mechanism does not quite reflect the relevant WZ
consistency condition: the latter should hold for full diffeomorphisms,
while the usual descent procedure relies only on the $GL(d)$ rotational
part of diffeomorphisms. Even though the resulting $GL(d)$ anomaly
descent yields the correct diffeomorphism anomaly \cite{Bardeen:1984pm},
it is not entirely transparent how this can be packaged into the BRST
algebra. For the case of the supersymmetry Ward identity and ``anomalous" terms
thereof, we find another deviation from the standard anomaly
descent, simply because there seems to be no place for the supersymmetry
parameters in the usual BRST algebra for anomaly descent.

The two share a common need for generalizing
the anomaly descent procedure.
We  should comment here that such a generalization of the BRST algebra
and the descent procedure thereof has been studied in the context of
supersymmetrized anomaly in the past. The most notable work is
Refs.~\cite{Kaiser:1988zg,Altevogt:1987fx} which inspired the bulk of
what we do here. Another such attempt was given later in Refs.~\cite{Baulieu:2006gx,Baulieu:2008id},
although their choice of the BRST operator and of the ghost differ from ours.
We should clarify that we focus on rigid supersymmetry and gauge/flavor anomalies here. See Refs.~\cite{Imbimbo:2018duh,Frob:2021sao}, e.g., for recent related discussions in the supergravity context.

Recall that the usual anomaly descent arise from a BRST algebra where we
replace
\bea
\tx d\;\rightarrow \; \tx d+\bs\ ,\qquad
\cA \;\rightarrow \;\Hat\cA\equiv \cA+v \ ,
\eea
with $\bs^2=0=(\tx d+\bs)^2$, so that
\be
\cF=\Hat \cF \equiv (\tx d +\bs)\Hat \cA+\Hat \cA^2\ .
\ee
Together they lead to
\be\label{poincareI&II}
P_{d+2}(\cf)=P_{d+2}(\Hat\cf)=(\tx d+\bs){\bf w}_{d+1}(\Hat\ca,\Hat\cf)=(\tx d+\bs){\bf w}_{d+1}(\Hat\ca,\cf)
\ee
for any given anomaly polynomial $P_{d+2}$.
As discussed in Section \ref{sec:background}, the right-most expression is now expanded in the ghost
number and equated to the left-most expression with no
ghost dependence, resulting in the standard descent formulae.

The generalized descent structure we are interested in arises
when, in addition to $\bs$, there exists an additional BRST odd
operator $\bc$ such that
\be
(\tx d+\bs+\bc)^2=0\ ,
\ee
as well. Of course, $\bs$ was meant to represent multiple
types of gauge transformation so the point of this additional
operator $\bc$ is that the action of $\bc$ on the connections and
the ghosts is not standard, i.e., as in $\bs v=-v^2$.
In our actual examples below, $\bc$
corresponds either to diffeomorphisms or to rigid supersymmetry. Note
that we do not necessarily demand that  $\bc^2=0$ or $(\bs+\bc)^2=0$
holds either, although they do hold when $\bc$ represents
diffeomorphims.

In order to generalize the BRST algebra in the presence
of $\bc$, we also add a new ghost $u$ and extend the BRST gauge field
further to $\hat\cA+u$. However, the relation between the $u$ ghost and
the operator $\bc$ would be rather different from that between $v$
and $\bs$. We need additional ghost to define $\bc$ but these do
not necessarily appear as $u$. In fact, for our two classes of examples in this Section,
we will take $u=0$ while $\bc$ and the ghost parameters thereof remain nontrivial.
Defining the BRST field strength
\be\label{generalized-fieldstrength}
\hat\cg\equiv(\tx d+{\bf s}+{\bf c})(\Hat\ca+u)+(\Hat\ca+u)^2,
\ee
where $\Hat\ca=\ca+v$ as before, this field strength
satisfies the generalized Bianchi identity
\be\label{generalized-Bianchi}
(\tx d+{\bf s}+{\bf c})\Hat\cg+(\Hat\ca+u)\Hat\cg-\Hat\cg(\Hat\ca+u)=0,
\ee
by virtue of the nilpotency of $\tx d+{\bf s}+{\bf c}$. We must
emphasize that we no longer have $\Hat\cg$ equal to $\cf$,
in view of how the $u$-ghost, or the absence thereof in examples below,
mismatches the BRST operator $\tx d+\bs+\bc$.

It follows that
\be
(\tx d+{\bf s}+{\bf c})P_{d+2}(\Hat\cg)=0,
\ee
and hence, locally
\be\label{poincareIII}
P_{d+2}(\Hat\cg)=(\tx d+{\bf s}+{\bf c}){\bf w}_{d+1}(\Hat\ca+u,\Hat\cg).
\ee
Subtracting either of the two relations in \eqref{poincareI&II} leads to the identity
\bebx\label{descentID}
(\tx d+{\bf s}+{\bf c}){\bf w}_{d+1}(\Hat\ca+u,\Hat\cg)-\tx d{\bf w}_{d+1}(\ca,\cf)=P_{d+2}(\Hat\cg)-P_{d+2}(\cf).
\eebx
The right hand side does not vanish, since $\Hat\cg\neq \cf$ in general. Instead, we find that
\bal\label{generalized-Russian}
\Hat\cG=&\;(\tx d+{\bf s}+{\bf c})(\Hat\ca+u)+(\Hat\ca+u)^2\NO\\
=&\;\Hat\cf+\tx du+({\bf s}+{\bf c})(\Hat\ca+u)+\Hat\ca u+u \Hat\ca+u^2\NO\\
=&\;\cf+{\bf c}\ca+{\bf s}u+\{v, u\}+\bc v+\bc u+{\tx d}_{\Hat \cA} u+u^2.
\eal
The difference $\Hat\cg-\cf$  depends on the operator $\bc$ and the choice of $u$.
Conversely, $u$ may be fixed by requiring the difference $\Hat\cg-\cf$ to be of a specific form.\footnote{As mentioned earlier, we shall take $u=0$ below, as it allows a single modification of the descent procedure to cover both diffeomorphisms and supersymmetry. Another natural choice for $u$ is discussed in Appendix \ref{sec:gendescent}.}

Given that $\Hat\cg-\cf$ is nonzero in general, we proceed by defining
\bebx\label{X}
P_{d+2}(\Hat\cg)-P_{d+2}(\cf)\equiv \sum_{k\geq 1} X^{(k)}_{d+2-k},
\eebx
where the integer $k$ again indicates the (generalized) ghost number.
The nonzero $X$'s can be viewed as obstructions to the standard anomaly descent
procedure.

As we shall see, in relevant examples, these $X$'s can be
themselves reconstructed by the action of  $(\bs+\bc)$ and $\tx d$ on
more elemental  quantities, to be denoted as $Y$'s and $Z$'s
respectively, provided that we remember that we want
a local functional on $d$ dimensional spacetime. This way, we
once again obtain a generalized anomaly annihilated by $(\bs+\bc)$.
The resulting anomaly would receive contributions from both
the left hand side and the  right hand side of (\ref{descentID}).

A simplest example of this, it turns out, is the familiar
diffeomorphism anomaly. Although we are accustomed to computing
the diffeomorphism anomaly via $GL(d)$ descent, we have reviewed
in Sec.~\ref{sec:diff} how the result actually obeys the
consistency condition of full diffeomorphisms. Next,
we will illustrate the above generalized anomaly descent for
this example and move on to supersymmetry later. In fact, our generalized descent procedure is inspired by an early attempt of constructing a supersymmetric descent procedure \cite{Kaiser:1988zg,Altevogt:1987fx,Altevogt:1989fw}.

\subsection{Diffeomorphism Anomaly Revisited}

Let us start by considering how the diffeomorphism
WZ consistency condition can be elevated to a BRST form. For
this we elevate $\cL_\xi$ to an operator $\bc$, with the unit
ghost number, such that
\bea\label{newBRST}
0={\tx d}^2=({\tx d} + \bs_g)^2=({\tx d} +\bs_g +\bc)^2\ ,
\eea
with $\bs_g$, for now, restricted to the internal gauge
transformation. The entire diffeomorphisms are carried
by $\bc$.

Given how $\cL_\xi = \cL'_\xi+\delta^{GL(d)}_{-\partial\xi}$
in general, more care is needed to define the action of $\bc$
on the Christoffel connections and the accompanying ghost.
On the connections
\bea
\bc A =\cL_\bx A =\cL'_\bx A\, \qquad \bc \Gamma =\cL_{\bx}'\Gamma+{\tx d}_ \Gamma(-\partial\bx)
\eea
while the action of $\bc$ on the ghosts is
\bea
\bc v_g =\cL_\bx v_g\ ,\qquad \bc\bx^\mu ={\bx}^\alpha \partial_\alpha \bx^\mu \ , \qquad \bs_g\bx=0
\eea
where $v_g$ is the gauge part of the ghost. It follows that
\bea
\bc (-\partial\bx) = - (-\partial\bx)^2 +\cL_\bx'(-\partial \bx)
\eea
where again $\cL'$ treats $(-\partial \bx)$ as if the latter is
a matrix-valued function. One can see that $\bc$ in part plays
the role of $\bs$ on the diffeomorphism sector but incorporates a
full diffeomorphism rather than $GL(d)$ only.

With this, $\bc^2=0$ by itself. For instance,
\bea
\bc^2v=\bc({\bx}^\alpha \partial_\alpha v) 
=(\bx^\alpha\partial_\alpha{\bx}^\mu) \partial_\mu v -
({\bx}^\alpha \partial_\alpha (\bx^\mu \partial_\mu v))
= -\bx^\alpha\bx^\mu\partial_\alpha\partial_\mu v=0
\eea
due to the Grassmannian properties of $\bx^\mu$. On tensors
(as well as on the Christoffel connection), it suffices to
consider $\bx=\chi \zeta+\chi' \xi$ with
a pair of Grassmannian coefficients, $\chi$ and $\chi'$, and a pair of
arbitrary vectors, $\zeta$ and $\xi$, whereby $ \bc(\bx^\mu) =\chi\chi'[\zeta,\xi]^\mu
$ holds so that
\bea
\bc\bc V = \cL_{\bc(\bx)}V -
\cL_{\chi\zeta+\chi'\xi}\left(\cL_{\chi\zeta+\chi'\xi}V\right)
=\chi\chi' \left(\cL_{[\zeta,\xi]}V-[\cL_\zeta,\cL_\xi]V \right)=0.
\eea
With $\bs_g$ being restricted solely to gauge transformations and $\bc$ representing
diffeomorphisms, $\bs_g$ and $\bc$ anticommute. Together with $\bc^2=0$,
this suffices to verify the nilpotency of the BRST operator \eqref{newBRST}  ${\tx d} +\bs_g +\bc$.

On the other hand, differently from $v_g$, the $\bx$ ghost  cannot  be naturally added
to  $\cA$, since it is really a vector, hence a directional
derivative. Instead we can define $v=v_g+(-\partial \bx)$ which includes all gauge rotations plus the $GL(d)$ rotation generated by diffeomorphisms.
Concurrently it is natural to split
\bea
\bc =\bs_{GL(d)}+\bc'
\eea
so that $\bc'$ retains the purely translational part, $\cL'_\bx$.
Its rotational part can be treated on an equal footing
with $\bs_g$ leading to the redefinition of the BRST operator
\bea
{\tx d}+\bs_g+\bc ={\tx d}+\bs+\bc'
\eea
with $\bs=\bs_g+\bs_{GL(d)}$. One may now take $v=v_g+(-\partial \bx)$
and since there is no other ghost that can be added
naturally, $u=0$. With such a mismatching ghost, we construct the BRST field
strength
\bea\label{G}
\Hat\cG\equiv ({\tx d} +\bs+ \bc')\Hat\cA + \Hat\cA^2= \Hat\cF+\bc'\Hat\cA   = \cF+\bc'(\cA+v)
\eea
which nevertheless obeys the same Bianchi identity as above
\bea
({\tx d} +\bs +\bc')\Hat\cG + \Hat\cA\Hat\cG-\Hat\cG\Hat\cA
\;=\; \bc' [\cF]-\bc'[({\tx d} +\bs)(\Hat\cA) +(\Hat\cA)^2]\;=\;0 .
\eea
We have invoked here $(\bc')^2+\{{\tx d} +\bs,\bc'\}=0$ that follows from (\ref{newBRST}).

An immediate consequence of the Bianchi identity is that an invariant polynomial $P_{d+2}(\Hat\cG) $ satisfies
\bea
&&({\tx d} +\bs +\bc')P_{d+2}(\Hat\cG)=0\ , \cr\cr
&&P_{d+2}(\Hat\cG)= (d+\bs +\bc'){\bf w}_{d+1}(\Hat\cA, \Hat\cG)\,.
\eea
The the two sides of the  second equation can be expanded respectively as
\bea
P_{d+2}(\Hat\cG)&=& P_{d+2}(\cF)+\sum_{k\ge 1} X^{(k)}_{d+2-k}(\bx,v,\cA,\cF) \cr\cr
{\bf w}_{d+1}(\Hat\cA, \Hat\cG) &= &{\bf w}_{d+1}(\cA,\cF) + \sum_{k\ge 1} W^{(k)}_{d+1-k}(\bx,v,\cA,\cF)
\eea
where $k$ keeps track of the net number of ghosts. A pair of descent towers now follows
\bea\label{2towers}
(\bs +\bc') X^{(k)}_{d+2-k} + {\tx d}X^{(k+1)}_{d+1-k} = 0 \ , \qquad   X^{(k+1)}_{d+1-k}= (\bs +\bc') W^{(k)}_{d+1-k} + {\tx d} W^{(k+1)}_{d-k}
\eea
with $X^{(0)}_{d+2}\equiv P_{d+2}(\cF)$ and $W^{(0)}_{d+1}\equiv {\bf w}_{d+1}(\cA,\cF)$.

Given $(\bs_g+\bc)^2=0$, the first equation in (\ref{2towers}) implies that
\bea
(\bs+\bc')\int X^{(k)}_{d+2-k} =0 \quad\rightarrow\quad \int X^{(k)}_{d+2-k} =(\bs+\bc')\int Y^{(k-1)}_{d+2-k}
\eea
for some $Y$'s that are not ${\tx d} $-exact. Taking $k=1$ in
(\ref{2towers}) we obtain
\bea
0=(\bs+\bc')\int_{M_d}  \left(W^{(1)}_{d} - Y^{(1)}_{d}\right),
\eea
a solution to the WZ consistency conditions,  which now due to  $\bs+\bc'=\bs_g+\bc$  are extended to include general diffeomorphisms $\delta_\xi$.

This may appear to be a tautology, since $Y^{(1)}=W^{(1)}$ is seemingly also acceptable, leading to a trivial solution to the BRST
version of the WZ consistency conditions. However, the point is that
a different nontrivial solution for $Y^{(1)}$ can be
found from
\bea\label{altX}
\sum_{k\ge 0} X^{(k+1)}_{d+1-k} &=& P_{d+2}(\Hat\cG)- P_{d+2}(\cF)\cr\cr
 &=& ({\tx d} +\bs+\bc')\sum_{l\ge 0} \left(W^{(l)}_{d+1-l}-{\bf w}^{(l)}_{d+1-l}\right) + \bc'\sum_{l\ge 0} {\bf w}^{(l)}_{d+1-l}\,.
\eea
In particular, the action of $\bc'$ on the $d$-form ${\bf w}^{(1)}_d$
is nothing  but $\cL_\bx'$, given that all $GL(d)$ indices are summed
over. Using the identity $\cL_\bx'(\cdots) =\bx\lrcorner {\tx d} (\cdots)
- {\tx d} (\bx\lrcorner\cdots)$ we find
\bea\label{X22}
X^{(2)}_d 
&=& (\bs+\bc') \left(W^{(1)}_d-{\bf w}_d^{(1)}\right) + {\tx d} \left(W^{(2)}_{d-1}-{\bf w}_{d-1}^{(2)}-  \bx\lrcorner  {\bf w}^{(1)}_d \right) +
\bx\lrcorner {\tx d} {\bf w}^{(1)}_d
\eea
which gives a different, nontrivial solution for $Y^{(1)}_d$.

We may  recall once again that ${\tx d} {\bf w}^{(1)}_d$ actually vanishes identically
once we restrict to the physical $d$ dimension and coordinates thereof, and the equation \eqref{X22} yields
\bea
0=(\bs+\bc')\int_{M_d}  \left(W^{(1)}_{d} - Y^{(1)}_{d}\right)=(\bs+\bc')\int_{M_d}  {\bf w}^{(1)}_{d}(v;\cA,\cF)
\eea
or more concretely
\bea
0&=&(\bs_g+\bc)\left(\int {\bf w}_d^{(1)}(-\partial\bx;\Gamma,R;F)
+\int {\bf w}_d^{(1)}(v_g;A,F;R)\right)
\eea
bringing us to the known solution, but via simpler BRST route
with diffeomorphisms now honestly incorporated.

\subsection{Anomaly Descent with Supersymmetry}

Now we can apply the same procedure to include supersymmetry
in the presence of gauge/flavor anomalies. For this, we use the
same $\bs=\bs_g$, $v=v_g$, and add
\bea
\bc=\bq+\mathbf a
\eea
where $\mathbf a$ acts universally as a translation by a
constant vector-valued ghost $a^\mu$,
\bea
\mathbf a (\cdots) =a^\mu\partial_\mu (\cdots) = \cL_a (\cdots)
\eea
while $\bq$ is the BRST version of the rigid supersymmetry.
Let us now use these in the generalized anomaly descent.

Unlike with the diffeomorphism case above, however, the
descent process remains universal only down to the first
term in the supersymmetry completion of the anomaly, which as
we have seen earlier can be expressed in terms of the BZ current. The rest, which is manifestly gauge invariant, depends very much on the spacetime dimension and amount of supersymmetry. We will therefore concentrate on outlining
the general procedure only up to this universal supersymmetry
completion. This way we can distill the previous attempts of constructing a supersymmetric descent formalism. The full form of the supersymmetry completion of the gauge/flavor anomaly in different cases can be found in Section \ref{sec:inflow} or in the literature.
See for example Refs.~\cite{Itoyama:1985qi,Guadagnini:1985ea,Itoyama:1985ni,Kaiser:1988zg,Altevogt:1987fx,Altevogt:1989fw}.

Let us consider an arbitrary even dimension $d$ and the minimal
supersymmetry thereof. For off-shell multiplets, the smallest possible field content arises for $d=6$ $\cn=(1,0)$, while for one-shell multiplets, one can go up to
$d=10$, $\cn=1$. As such the fermions would obey a Majorana
condition of some kind. The action of $\bq$ involves
the c-number-valued Majorana spinor ghost $\a$ and takes the form
\bea
\bq A_\m&=&\;-\tfrac12\bar\a\g_\m\l,\NO\\
\bq \l&=&\;\tfrac14\g^{\m\n}F_{\m\n}^a+\cdots\NO\\
\bq v&=&\;\tfrac14\bar\a\g^\m\a A_\m,\NO\\
\bq \a &=&0,\NO\\
&\cdots& \NO\\
\bq a^\m&=&\;-\tfrac14\bar\a\g^\m\a
\eea
where the ellipses denote transformations involving auxiliary
fields for off-shell transformations, e.g., as in \eqref{4d-N=2-susy-transformations}. Let us also note that
the translation operator $\mathbf a$ acting on $a$
and on $\alpha$ vanishes as these ghosts are taken to
be constant.\footnote{Elevating these to position-dependent quantities must entail couplings to external supergravity and is needed if we wish to extend this
to the cases with a diffeomorphism anomaly. We believe that
the same supersymmetric descent mechanism can be extended to that case as well.}

Choosing $u=0$, the generalized descent for $\bc$  works precisely the same way as already outlined.  In particular, for rigid supersymmetry
\bea
X^{(2)}_d &=& (\bs+\bc) \left(W^{(1)}_d-{\bf w}_d^{(1)}\right) + {\tx d} \left(W^{(2)}_{d-1}-{\bf w}_{d-1}^{(2)} \right) +
\bc {\bf w}^{(1)}_d
\eea
where in the last term ${\bf w}_d^{(1)}$ depends neither on $\bc$ ghost nor
on the superpartner. The $\bc$ operation on it can be decomposed into
three parts; one is the SUSY transformation $\bq$ acting on $A$ and $F$,
the other also $\bq$ acting on $v$, and finally the action of $\mathbf a$
as $a$-gradient on $v,A, F$. In particular the last $\mathbf a$ acting
on the differential $d$-form produces total derivative terms, which will
eventually vanish upon $M_d$ integration.

The first part can be written, using the same anti-derivative $l$, as
\bea
\delta_{\bq A} =-{\tx d} l_{\bq A}+l_{\bq A}{\tx d}\,.
\eea
The sign flip (cf. eq. \eqref{Delta}) is due to the fact that $l_{\bq A}$ not only carries a single ghost number but also reduces the rank of the differential form, and hence is BRST even.
Modulo terms that drop out upon  $M_d \,$-integration,
\bea
&&\bc {\bf w}^{(1)}_d \;=\; - l_{\bq A}\left((\bs {\bf w}^{(0)}_{d+1}(A,F)\right)
+ {\bf w}^{(1)}_d(\bq v;A,F) +\cdots\cr\cr &=&(\bs+\bc)\left( - l_{\bq A} {\bf w}^{(0)}_{d+1}(A,F)\right) +  \bc \left(l_{\bq A} {\bf w}^{(0)}_{d+1}(A,F)\right)   + {\bf w}^{(1)}_d(\bq v;A,F) +\cdots
\eea
since ${\bf w}^{(1)}_d$ is linear in $v$. Thus, we arrive at
\bea
X^{(2)}_d &=& (\bs+\bc) \left(W^{(1)}_d-{\bf w}_d^{(1)} -  l_{\bq A} {\bf w}^{(0)}_{d+1}\right)     \cr\cr
&& + \; \bq \left(l_{\bq A} {\bf w}^{(0)}_{d+1}(A,F)\right)   + {\bf w}^{(1)}_d(\bq v;A,F) +\cdots
\eea
where we again used the fact that $\mathbf a$ produces a total derivative term
at most. Note that the last term in the first line is
\bea
\int l_{\bq A} {\bf w}^{(0)}_{d+1}=\bq A\cdot X(A,F)
\eea
where $X$ is the BZ current. This is exactly the
$\lambda$-linear term that we would have found by imposing
the mixed SUSY-gauge WZ consistency condition as in Refs.~\cite{Itoyama:1985qi,Guadagnini:1985ea,Itoyama:1985ni}.

The question is then: Does the second line produce a term
which has the form $(\bs+\bc)$ acting on something else?
A useful middle step is to separate out those terms with
no gaugino attached by rewriting,
\bea
(\bq A)\cdot \bq X(A,F)  + \left((\bq\bq A)\cdot X(A,F)   + {\bf w}^{(1)}_d(\bq v;A,F)\right)\,.
\eea
Now, we know from the above that
\bea
\bq v= \xi\lrcorner A ,\qquad \xi^\m\equiv \tfrac14\bar\a\g^\m\a
\eea
while
\bea
\bq\bq A_\rho = \bq\left(-\tfrac12\bar\a\g_\rho\l \right)
=\frac{1}{16}\bar\a[\g_\rho, \g^{\m\n}]\a F_{\m\n} =-(\xi\lrcorner F)_\rho
\eea
where we use $\bar\a\g_{\rho\m\nu}\a=0$ for c-valued Majorana $\a$.\footnote{The
transformation of $\lambda$ into the auxiliary field can
potentially spoil this; however, one can see that the Majorana
nature of $\alpha$ is such that this piece vanishes identically,
for the same reason that $\bar\a\g_{\rho\m\nu}\a=0$.} Combining the two, we find
\bea\label{Itoyama1}
\left((\bq\bq A)\cdot X(A,F)   + {\bf w}^{(1)}_d(\bq v;A,F)\right)=-(\xi\lrcorner F)\cdot X - (\xi\lrcorner A) \cdot\langle \nabla\cdot J\rangle =0
\eea
for exactly the same reason how the covariant gauge current,
rather than the consistent current, had to enter in the
diffeomorphism Ward identity as in Section \ref{sec:background}.

As such, we are left with a single term that can further
contribute to $Y^{(1)}$. Its form from Section \ref{sec:diff} suggests
that the answer must be gauge-invariant so we hope to find $\Delta Y^{(1)}_{\rm inv}$
such that
\bea
(\bq A)\cdot \bq X(A,F) = \int (\bs+\bc)\,\Delta Y^{(1)}_{\rm inv}  = \int \bq\Delta Y^{(1)}_{\rm inv}
\eea
where again, $\mathbf a$ acting on $\Delta Y^{(1)}_{\rm inv}$ turns into
a total derivative. This, together with (\ref{Itoyama1}), would precisely
elevate the solutions of \cite{Itoyama:1985qi,Guadagnini:1985ea,Itoyama:1985ni} to the SUSY-SUSY WZ consistency condition
to the BRST level in our generalized descent context.

For this to work, one necessary condition is that the left-hand-side
is itself invariant under gauge transformations. In fact, we claim that
this pattern of $\bq A\cdot \bq X$ being
gauge-invariant is universal, even though the BZ current
$X$ is not gauge covariant. One can see this formally via the
very definition of the BZ current,
\bea
X=J_{\rm cov}- J = J_{\rm cov}-\frac{\delta W}{\delta A}
\eea
whereby
\bea
\bq A\cdot \bq X = \bq A\cdot \bq J_{\rm cov}-(\bq A)(\bq A)\cdot \frac{\delta^2 W}{\delta A\delta A}\,.
\eea
The first piece is manifestly invariant with $\bq A$ being covariant,
while the second, potentially noninvariant piece vanishes since,
component-wise, $\bq A_\mu$ is BRST odd while, $A_\mu$ is BRST even.

Given the anomaly polynomial $P_{d+2}(F)$, a sum of products of
symmetrized traces of its $d/2+1$ arguments as usual, what would
be the explicit expression for $\bq A\cdot \bq X$? We start from
the usual transgression formula for $\mathbf w_{d+1}^{(0)}$ and
arrive at
\bea
\bq A\cdot X = (d/2)(d/2+1)\int_0^1 dt \int_{M_d} \,P_{d+2} (A, t\bq A, F_t,F_t,\cdots, F_t)
\eea
with $A_t\equiv tA$ and $F_t\equiv dA_t+A_t^2=tF+(t^2-t)A^2$.
The integrand of $\bq A\cdot \bq X$ becomes
\bea
&&P_{d+2} (\bq A, t\bq A, F_t,\cdots, F_t) + (d/2-1)  P_{d+2} (A, t\bq A, {\tx d}_{tA}(t\bq A) ,F_t,\cdots, F_t) \cr\cr
&=&\frac{1}{2} \partial_t P_{d+2} (t\bq A, t\bq A, F_t ,F_t,\cdots, F_t)
+\cdots
\eea
which gives, dropping the $\tx d$-exact term in the ellipses,
\bea\label{qAqX}
\bq A\cdot \bq X = \frac{(d/2)(d/2+1)}{2}\int_{M_d}P_{d+2} (\bq A, \bq A, F,F,\cdots, F)
\eea
with $d/2-1$ $F$'s inside.

In short, we have shown that the supersymmetry completion of the anomaly
has the universal decomposition
\bea
\int\left({\bf w}_d^{(1)} + l_{\bq A} {\bf w}^{(0)}_{d+1} - \Delta Y^{(1)}_{\rm inv}\right)
\eea
that is, the standard anomaly, the BZ current contracted against $\bq A$,
and the last, invariant part of the supersymmetry completion that solves
\bea
\bq \int \Delta Y^{(1)}_{\rm inv} = \frac{(d/2)(d/2+1)}{2}\int_{M_d}P_{d+2} (\bq A, \bq A, F,F,\cdots, F)
\eea
where $P_{d+2}$ is the anomaly polynomial responsible for $\mathbf w^{(k)}_{d+1-k}$'s
via the standard anomaly descent.

As we warned earlier, the precise solution for $\Delta Y^{(1)}_{\rm inv}$ strongly depends on spacetime dimensions, number of supersymmetries, and on-shell vs. off-shell.
For $d\le 4$, we have already given several answers via Chern-Simons anomaly
inflow. For examples in higher dimensions, such as $d=10$ and $\cn=1$ or $d=6$
and $\cn=(1,0)$, we refer the readers to Ref.~\cite{Kaiser:1988zg,Altevogt:1987fx,Altevogt:1989fw}, which
in fact inspired  the general descent procedure we gave in the section.

Even though we started with minimal supersymmetry in even $d$,
the actual spacetime dimension is flexible. In other words, the same
descent mechanism works as long as the theory comes from dimensional reduction
and truncation of some higher $d'>d$ theory with minimal supersymmetry;
we only need to make sure that the $\gamma$ matrices above are those of the
original $d'$ dimensional Majorana spinor. For example, one can make use
of the $d=6$ $\cn=(1,0)$ example for $d=4$ $\cN=2$, although in this latter
case, the chirality is lost and gauge/flavor anomaly would be absent.
A more interesting example would come about by a further truncation of supersymmetry as we reduce $d'\rightarrow d$. In fact the same
sort of idea was employed in the previous section where, for example,
we start with anomaly inflow from a single $d=3$ $\cn=3$ Chern-Simons
to generate the anomaly of $d=2$ $\cn=(p,q)$ for various $p,q\le 3$.

Finally, we should mention again that the supersymmetry completion of
anomalies has been the subject of past studies, and much of what we
outlined in this section has already appeared  in bits and pieces. For
instance, an equivalent form of (\ref{qAqX}) has appeared
as early as in Ref.~\cite{Itoyama:1985qi,Itoyama:1985ni}, albeit as
an on-shell statement and without a BRST formulation. The present discussion
combines and organizes these earlier works in a single coherent framework,
where the central role is played by  the Bianchi identity (\ref{generalized-Bianchi}) and the generalized Russian formula (\ref{generalized-Russian}).

\subsection{Supersymmetric anomaly inflow from anomaly descent}

A direct connection between the supersymmetric anomaly inflow mechanism discussed in Section \ref{sec:inflow} and the generalized anomaly descent exists if and only if $X^{(1)}_{d+1}$ satisfies the normality condition \eqref{normality}, i.e.
\be\label{X1-normality}
X^{(1)}_{d+1}=({\bf s}+{\bf c})Y^{(0)}_{d+1}+\tx dZ^{(1)}_{d},
\ee
for some nontrivial $Y^{(0)}_{d+1}\neq  W^{(0)}_{d+1}$. If $X^{(1)}_{d+1}$ can be expressed in this form,  then the second descent equation in \eqref{X&W} determines that
\be
(\bs +\bc) \big(W^{(0)}_{d+1}-Y^{(0)}_{d+1}\big)=\tx d\big(Z^{(1)}_{d}- W^{(1)}_{d}\big),
\ee
which allows us to identify $W^{(0)}_{d+1}-Y^{(0)}_{d+1}$ with a supersymmetric Chern-Simons form and $Z^{(1)}_{d}- W^{(1)}_{d}$ with the nontrivial solution of the WZ consistency conditions.

In order to verify that $Z^{(1)}_{d}- W^{(1)}_{d}$ coincides (up to $(\bs +\bc)$-exact terms, i.e. local counterterms) with the solution of the WZ conditions obtained earlier in this section we observe that the first descent equation in \eqref{X&W} implies that
\be
\tx d\big((\bs +\bc)Z^{(1)}_{d}-X^{(2)}_{d}\big) = 0,
\ee
and hence,
\be
X^{(2)}_{d}=(\bs +\bc)Z^{(1)}_{d}+\tx d Z^{(2)}_{d-1}.
\ee
We recognize this as the normality condition on $X^{(2)}_{d}$ we saw above. In particular, we identify
\be
Z^{(1)}_{d}=Y^{(1)}_{d},
\ee
up to $(\bs +\bc)$-exact terms. This confirms that the nontrivial solution of the WZ consistency conditions can be obtained from the Chern-Simons form $W^{(0)}_{d+1}-Y^{(0)}_{d+1}$ and takes the form
\be
\text{anomaly} = Z^{(1)}_{d}-W^{(1)}_{d} = Y^{(1)}_{d}-W^{(1)}_{d}+ (\bs +\bc)\text{-exact terms}.
\ee

To illustrate this connection between supersymmetric anomaly inflow and generalized descent, let us consider a simple example from Section \ref{sec:inflow}. From the supersymmetric anomaly inflow analysis in Section \ref{sec:inflow} we expect that $X_{d+1}^{(1)}$ can be expressed in normal form for all multiplets in $d$ dimensions that can be obtained by dimensional reduction from $d+1$ dimensions, without any further truncation. The 4d $\cn=1$ flavor anomaly is therefore not a suitable example, since it must first be embedded in the $\cn=2$ flavor multiplet in order to be obtainable from a 5d supersymmetric Chern-Simons action. The simplest example where $X_{d+1}^{(1)}$ can be expressed in normal form is the $\cn=(1,1)$ flavor anomaly in 2d, and the corresponding $\cn=1$ supersymmetric Chern-Simons action in three dimensions.

The $\cn=1$ vector multiplet in three dimensions can be obtained by consistent truncation of the $\cn=3$ multiplet, as we discussed in Section \ref{sec:inflow}. Changing to anti Hermitian generators, $t^a_{\rm h.}\to i t_{\rm a.h.}^a$, and scaling the gaugino and the supersymmetry parameter as
\be
\l^a\to i \l^a,\qquad \e\to \frac i2\e,
\ee
in order to match the conventions of the present section, the (canonically normalized) $\cn=1$ supersymmetric Chern-Simons form is given by
\be\label{d=3-N=1-CS}
\Om_{\rm CS}=\tr\Big(A\tx dA+\frac23 A^3-\bar\l\l\ast \bb1\Big),
\ee
where we have fixed the choice of convention \eqref{3d-gamma-convention} by setting $\k=-1$.

The $\cn=1$ supersymmetry transformations become
\bal
\label{susy-N=1-vector}
\d_Q(\e) A_\m=&\;-\tfrac12\bar\e\g_\m\l,\NO\\
\d_Q(\e)\l_\m=&\;\tfrac{1}{4}(\g^{\r\s}\e)_\m F_{\r\s},
\eal
while the complete form of the corresponding BRST transformations is
\begin{align}
\begin{aligned}
\bs A_\m=&\;\cd_\m v,\\
\bs \l=&\;\{\l,v\},\\
\bs v=&-v^2,\\
\bs \a=&\;0,
\end{aligned}\hspace{1.cm}
\begin{aligned}
\bc A_\m=&\;-\tfrac12\bar\a\g_\m\l^a+a^\n\pa_\n A_\m,\\
\bc \l=&\;\tfrac14\g^{\m\n}F_{\m\n}^a\a+a^\n\pa_\n\l,\\
\bc v=&\;\tfrac14\bar\a\g^\m\a A_\m+a^\n\pa_\n v,\\
\bc a^\m=&\;-\tfrac14\bar\a\g^\m\a.
\end{aligned}
\end{align}

From the definition of the generalized descent variables $X_{d+2-k}^{(k)}$ in \eqref{X} follows that
\bal
X_{d+1}^{(1)}=&\;\tfrac{d+2}{2}P_{d+2}\big(\bc A,F^{\tfrac d2}\big).
\eal
In particular, for $d=2$ we have
\bal
X_{3}^{(1)}=&\;2\tr(\bc A F)\NO\\
=&\;2\tr\big((-\tfrac12\bar\a\g\l+\tx d i_a A+i_a\tx d A) F\big)\NO\\
=&\;2\tr\big(-\tfrac12\bar\a\g\l F+\tx d (i_a A) F+i_a(F-A^2) F\big)\NO\\
=&\;\tr\big(-\bar\a\g\l F+2\tx d_A (i_a A) F+i_a F^2\big)\NO\\
=&\;\tr(-\bar\a\g\l F)+\tx d\,\tr(2i_a A F),
\eal
where in the last line we have used the fact that $F^2=0$ in three dimensions.

Moreover, using the BRST transformation of $\l$ we obtain
\be
{\bf c}\,\tr(\bar\l\l)=\frac12\ve^{\m\n\r}\tr\big(\bar\a\g_\r\l F_{\m\n}\big)+\pa_\m\tr\big(a^\m\bar\l\l\big),
\ee
or in form notation
\be
{\bf c}\,\tr(\bar\l\l\ast \bb1)=-\tr(\bar\a\g\l F)-\tx d\,\tr(\ast a\bar\l\l).
\ee
We therefore conclude that $X_{3}^{(1)}$ can be expressed in normal form as
\be
X_{3}^{(1)}=({\bf s}+{\bf c})\tr(\bar\l\l\ast \bb1)+\tx d\,\tr(2i_a A F-\ast a\bar\l\l),
\ee
from which we read off
\be
Y^{(0)}_{3}=\tr(\bar\l\l\ast \bb1),\qquad Z^{(1)}_{2}=\tr(2i_a A F-\ast a\bar\l\l).
\ee
Hence, the supersymmetric Chern-Simons form \eqref{d=3-N=1-CS} is given by
\be
\O_{\rm CS}=W_3^{(0)}-Y_3^{(0)},
\ee
in agreement with the general argument provided above.

\section*{Acknowledgments}

We thank H. Itoyama for communication on his pioneering work. We also acknowledge use of the symbolic computation package FieldsX \cite{Frob:2020gdh} and IP would like to thank Markus Fr\"ob for helpful correspondence and early access to an updated version of the package. The work of RM is supported in part by ERC grants 772408-Stringlandscape and 787320-QBH Structure. The work of IP and PY is supported, respectively, by the KIAS Individual Grants PG064402 and PG005704 at the Korea Institute for Advanced Study.

\appendix

\renewcommand{\thesection}{\Alph{section}}
\renewcommand{\theequation}{\Alph{section}.\arabic{equation}}

\section*{Appendix}
\setcounter{section}{0}

\section{Additional Notions}
\label{sec:gendescent}

In Section \ref{sec:descent} we have tried to simplify the discussion and avoid introducing new notions and definitions. However, as mentioned,  our generalized descent procedure is inspired by earlier work on supersymmetric descent  \cite{Kaiser:1988zg,Altevogt:1987fx,Altevogt:1989fw,Baulieu:2006gx,Baulieu:2008id}.  We collect here some definitions and ideas that have appeared in some form in the earlier literature and have been kept implicit in our presentation.

Without splitting the $\bc$ part of the BRST operator ${\tx d} +\bs +\bc$ into parts and reserving $\bs$ for the gauge transformations only, we can still recover the general pair of the descent towers for $X$ and $W$ outlined in Section \ref{sec:descent}:
\bal\label{X&W}
&(\bs +\bc) X^{(k)}_{d+2-k} + {\tx d} X^{(k+1)}_{d+1-k} = 0, \NO\\
& X^{(k+1)}_{d+1-k}= (\bs +\bc) W^{(k)}_{d+1-k} + {\tx d} W^{(k+1)}_{d-k},
\eal
which can be  subject to further (non-universal) refinements.

\paragraph{Filtration} The descent relations we have discussed so far hold for any $\bc$, as long as $(\tx d+\bs+\bc)^2=0$. However, this structure is enriched when $\bc$ possesses certain additional properties. One such case is when $\bs+\bc$ is itself nilpotent, i.e.
\be\label{filtration}
\bs^2=0,\qquad (\bs+\bc)^2=0,
\ee
where generically $\bc^2\neq 0$, and both $\bs$ and $\bc$ anticommute with the exterior derivative $\tx d$. This structure is formally known as a filtration  of the BRST cohomology \cite{Dixon:1991wi,Piguet:1995er} and it is practically very useful for computing the cohomology of $\bs+\bc$, given that of the simpler BRST operator $\bs$. For example, as we discuss in Section \ref{sec:diff} (see also \cite{Bertlmann:1996xk} and references therein), the gravitational anomaly descent can be formulated in terms of the simpler descent for a $GL(d)$ gauge symmetry. More generally, the structure \eqref{filtration} ensures that the cohomology of $\bs+\bc$ is a subspace of the cohomology of $\bs$ (see Proposition 5.6 in \cite{Piguet:1995er}). The same holds for the cohomologies modulo $\tx d$.

\paragraph{Grading refinement} It is often useful to further decompose the coefficients $X^{(k)}_{d+2-k}$ and  $W^{(k)}_{d+1-k}$ in \eqref{X&W}, for example, according to different types of ghosts. In particular, we may distinguish between $\bs$ and $\bc$ ghosts by writing
\be\label{refinement}
W^{(k)}_{d+1-k}=\sum_{l=0}^kW^{(l,k-l)}_{d+1-k},\qquad X^{(k)}_{d+2-k}=\sum_{l=0}^k X^{(l,k-l)}_{d+2-k},
\ee
where the first superscript corresponds to ${\bf s}$ ghosts and the second to ${\bf c}$ ghosts. However, such a decomposition is not always possible. This is the case e.g. when $\bs$ and $\bc$ possess the nilpotency properties \eqref{filtration}, which imply that
\be
\bs\bc+\bc\bs+\bc^2=0.
\ee
It follows that, in this case, unless the stronger conditions
\be
\bs\bc+\bc\bs=0,\qquad \bc^2=0,
\ee
hold, the $\bs$ and $\bc$ ghost numbers are not separately conserved.

When a grading refinement of the form \eqref{refinement} is possible, each descent relation in \eqref{X&W} splits accordingly into a set of equations with definite $(p,q)$ ghost number. For example, for $k=1$ the first equation in  \eqref{X&W} splits into the three relations:
\be
\tx d X_{d}^{(2,0)}+{\bf s}X^{(1,0)}_{d+1}=0,\qquad
\tx d X_{d}^{(1,1)}+{\bf c}X^{(1,0)}_{d+1}=0,\qquad
\tx d X_{d}^{(0,2)}=0,
\ee
since $X^{(0,1)}_{d+1}=0$. Similarly, the second equation in \eqref{X&W} for $k=1$ splits into the relations
\bal
&{\bf s}W^{(1,0)}_{d}+\tx dW^{(2,0)}_{d-1}=0,\NO\\
&{\bf s}W^{(0,1)}_{d}+{\bf c}W^{(1,0)}_{d}+\tx dW^{(1,1)}_{d-1}=-X^{(1,1)}_{d},\NO\\
&{\bf c}W^{(0,1)}_{d}+\tx dW^{(0,2)}_{d-1}=-X^{(0,2)}_{d}.
\eal

As we discuss next, a grading refinement of the form \eqref{refinement} is particularly useful when the coefficients $X^{(k)}_{d+2-k}$ satisfy two additional properties, namely ``normality" and ``orthogonality", which we now define.

\paragraph{Normality} Following \cite{Kaiser:1988zg}, we say that $X^{(k)}_{d+2-k}$ is ``normal'' if it takes the form
\be\label{normality}
X^{(k)}_{d+2-k}=({\bf s}+{\bf c})Y^{(k-1)}_{d+2-k}+\tx dZ^{(k)}_{d+1-k},
\ee
for some $Y^{(k-1)}_{d+2-k}$ and $Z^{(k)}_{d+1-k}$. Since the coefficients $X^{(k)}_{d+2-k}$ satisfy the descent relations \eqref{X&W}, this may seem like a tautology at first sight. However, the statement is not trivial in that one demands that $Y^{(k-1)}_{d+2-k}$ and $Z^{(k)}_{d+1-k}$ have different ghost content than respectively $W^{(k-1)}_{d+2-k}$ and $W^{(k)}_{d+1-k}$. This distinction becomes manifest when a grading refinement of the form \eqref{refinement} is possible.

It is useful to notice that if the normality condition \eqref{normality} holds for $X^{(k)}_{d+2-k}$, then it also holds for all $X^{(l)}_{d+2-l}$ with $l>k$. This can be proven by induction, showing first that it holds for $l=k+1$. From \eqref{X&W} follows that
\bal
0=\tx dX^{(k+1)}_{d+1-k}+({\bf s}+{\bf c})X^{(k)}_{d+2-k}=\tx d\big(X^{(k+1)}_{d+1-k}-({\bf s}+{\bf c})Z^{(k)}_{d+1-k}\big),
\eal
and so
\be
X^{(k+1)}_{d+1-k}=({\bf s}+{\bf c})Y^{(k)}_{d+1-k}+\tx dZ^{(k+1)}_{d-k},
\ee
with $Y^{(k)}_{d+1-k}=Z^{(k)}_{d+1-k}$. Hence, $X^{(k+1)}_{d+1-k}$ is also normal. This completes the proof.

\paragraph{Orthogonality} We have seen that the Russian formula \eqref{Russian} ensures that the coefficient $X^{(1)}_{d+1}$ contains only ghosts associated with the operator $\bc$. If this property extends to all higher order coefficients, we say that $X^{(k+1)}_{d+1-k}$ are ``orthogonal". Clearly, a prerequisite for this property is that a grading refinement of the form \eqref{refinement} exists. The orthogonality condition can be stated as
\be
X^{(l,k-l)}_{d+2-k}=0,\quad \forall l\neq 0,
\ee
or equivalently
\be\label{orthogonality}
X^{(k)}_{d+2-k}=X^{(0,k)}_{d+2-k}.
\ee
From the definition \eqref{X} of the coefficients $X^{(k)}_{d+2-k}$ follows that this property is guaranteed provided the difference $\Hat\cg -\cf$ between the field strengths does not depend on the gauge ghost $v$. This condition can be used as a possible criterion for the choice of the shift $u$ in the generalized field strength \eqref{generalized-fieldstrength}. Notice that if the coefficients $X^{(k)}_{d+2-k}$ are also normal, then the definition \eqref{normality} requires that
\be\label{refined-normality}
Y^{(k-1)}_{d+2-k}=Y^{(0,k-1)}_{d+2-k},\qquad {\bf s}Y^{(k-1)}_{d+2-k}=0,\qquad Z^{(k)}_{d+1-k}=Z^{(0,k)}_{d+1-k}.
\ee
In the main part of the paper we have opted for the choice $u=0$ in order to ensure the uniform description of diffeomorphisms and supersymmetry.

\bibliographystyle{JHEP}
\bibliography{references}

\end{document}

%% file: macros.tex
\def\){\right)}
\def\({\left( }
\def\]{\right] }
\def\[{\left[ }

\def\NO{\nonumber}

\newcommand{\be}{\begin{equation}}
\newcommand{\ee}{\end{equation}}

\def\bea{\begin{eqnarray}}
\def\eea{\end{eqnarray}}

\def\bal#1\eal{\begin{align}#1\end{align}}

\def\bald{\begin{aligned}}
\def\eald{\end{aligned}}

\def\bsub{\begin{subequations}}
\def\esub{\end{subequations}}

\def\beqx{\begin{displaymath}}
\def\eeqx{\end{displaymath}}

\newcommand{\bmat}{\left(\begin{array}}
\newcommand{\emat}{\end{array}\right)}

\def\a{\alpha}

\def\c{\chi}
\def\d{\delta}
\def\e{\epsilon}
\def\f{\phi}
\def\g{\gamma}
\def\h{\eta}
\def\j{\psi}
\def\k{\kappa}
\def\l{\lambda}
\def\m{\mu}
\def\n{\nu}
\def\o{\omega}
    \def\om{\omega}
\def\p{\pi}

    \def\th{\theta}
\def\r{\rho}
\def\s{\sigma}
\def\t{\tau}
\def\x{\xi}

\def\L{\Lambda}
\def\O{\Omega}
    \def\Om{\Omega}

\def\ve{\varepsilon}

    \def\vth{\vartheta}

\def\ca{{\cal A}}

\def\cd{{\cal D}}

\def\cf{{\cal F}}
\def\cg{{\cal G}}

\def\cl{{\cal L}}

\def\cn{{\cal N}}

\def\bb#1{\ensuremath{\mathbb{#1}}} 

\def\bo{{\raise-.3ex\hbox{\large$\Box$}}}               
\def\pa{\partial}                                       
\def\face{{\raise.2ex\hbox{$\displaystyle \bigodot$}\mskip-2.2mu \llap {$\ddot
        \smile$}}}                                   
\def\>{\rangle}                                      
\def\<{\langle}                                      

\def\tx#1{\text{#1}}
\def\slash#1{\rlap{\hbox{$\mskip 1 mu /$}}#1}        
\def\Hat#1{\widehat{#1}}                             
\def\leftrightarrowfill{$\mathsurround=0pt \mathord\leftarrow \mkern-6mu
        \cleaders\hbox{$\mkern-2mu \mathord- \mkern-2mu$}\hfill
        \mkern-6mu \mathord\rightarrow$}        
\def\dvec#1{\vbox{\ialign{##\crcr
        \leftrightarrowfill\crcr\noalign{\kern-1pt\nointerlineskip}
        $\hfil\displaystyle{#1}\hfil$\crcr}}}           
\def\tr{{\rm tr \,}}                                    

\def\-{\hphantom{-}}

